\documentclass[aps,twocolumn,pra,superscriptaddress,amsmath,amssymb]{revtex4-1}
\usepackage{dcolumn}
\usepackage{bm}
\usepackage{amsmath}
\usepackage{txfonts}
\usepackage[T1]{fontenc}
\usepackage{xspace}
\usepackage{ulem}
\usepackage{comment}
\usepackage{braket}
\usepackage{mhchem}
\usepackage{color}
\usepackage{ulem}
\newcommand{\means}[1]{\langle#1\rangle}

\ifx\pdfoutput\undefined
\usepackage[dvipdfmx]{graphicx}
\usepackage[dvipdfmx]{hyperref}
\usepackage[dvipdfmx]{color}
\usepackage[dvipdfmx]{xcolor}
\else
\usepackage{graphicx}
\usepackage{hyperref}
\usepackage{color}
\usepackage{xcolor}
\fi
\hypersetup{
        colorlinks=true,
        citecolor=blue,
        urlcolor=blue,
        linkcolor=blue
}

\begin{document}
\let\emph\textit

\title{Effects of magnetic fields and orbital angular momentum on excitonic condensation in two-orbital Hubbard model}

\author{Ryota Koga}
\author{Joji Nasu}
\affiliation{
  Department of Physics, Tohoku University, Sendai, Miyagi 980-8578, Japan
}

\date{\today}
\begin{abstract}
We investigate the magnetic-field effects on a two-orbital Hubbard model that describes multiple spin states.
The spin-state degrees of freedom in perovskite-type cobalt oxides have been explored due to their characteristic nature, where low-spin, intermediate-spin, and high-spin states play a crucial role in their magnetic properties caused by the competition between Hund coupling and crystalline field effects. 
Recent findings have suggested that this interplay leads to quantum mechanical hybridization of these spin states, which is interpreted as excitonic condensation.
When the ground state comprises primarily low-spin states because of dominant crystalline fields, fluctuations between different spin states are anticipated to arise from Zeeman splittings in higher-spin states. 
This offers an alternative approach to induce excitonic condensation using an external magnetic field.
To understand magnetic-field effects on excitonic condensation in multi-orbital systems, it is crucial to account for contributions from both spin and orbital degrees of freedom to magnetic properties.
Here, we study field-induced phenomena in the two-orbital Hubbard model by focusing on the role of the orbital angular momentum.
We comprehensively analyze this model on a square lattice employing the Hartree-Fock approximation.
Omitting contributions from the orbital moment, we find that an applied magnetic field gives rise to two excitonic phases, besides the spin-state ordered phase, between the nonmagnetic low-spin and spin-polarized high-spin phases. 
One of these excitonic phases manifests a staggered-type spin-state order, interpreted as an excitonic supersolid state.
Conversely, the other phase is not accompanied by it and exhibits only a spin polarization due to the applied magnetic field.
When spin-orbit coupling is present, this phase displays a ferrimagnetic spin alignment attributed to spin anisotropy. 
Our analysis also reveals that incorporating the contribution of the orbital magnetic moment to the Zeeman term significantly alters the overall structure of the phase diagram.
Notably, the orbital magnetization destabilizes the excitonic phase in contrast to scenarios without this contribution.
We also discuss the relevance of our findings to real materials, such as cobalt oxides.
\end{abstract}
\maketitle

%%%%%%%%%%%
\section{Introduction}

Excitonic insulating states, proposed nearly half a century ago, are characterized by the spontaneous condensation of excitons, pairs of electrons in conduction bands and holes in valence bands~\cite{Mott1961,Jerome1967,Halperin1968}.
% The appearance of this state was originally anticipated to emerge in semimetals or narrow-gap semiconductors when the exciton binding energy exceeds the band gap.
Since an exciton is a pair of particles with an $S=1/2$ spin, excitonic condensation is roughly classified into two types: spin-triplet and spin-singlet.
Recent studies on excitonic condensation primarily focused on the latter.
Amongst others, the transition metal chalcogenide $\ce{Ta2NiSe5}$ has been intensively investigated as a candidate of excitonic insulators~\cite{Wakisaka2009, Seki2014}.
This material undergoes a structural phase transition at $T_S \simeq 328$K~\cite{Salvo1986}.
Angle-resolved photoemission spectroscopy experiments have observed a flattening of the valence band below this temperature~\cite{Wakisaka2009, Seki2014}.
This band structure deformation is considered a manifestation of spin-singlet excitonic condensation associated with lattice distortion~\cite{Kaneko2013, Kaneko2015}.

Spin-triplet excitonic condensation is also predicted to manifest in real materials due to Hund coupling interaction rather than electron-lattice interactions.
Perovskite-type cobalt oxides have attracted considerable attention as candidate materials exhibiting spin-triplet excitonic condensation~\cite{Kunes2014condensation,Kunes2014instability}.
These materials, typified by $\ce{LaCoO3}$, have been studied as strongly correlated electron systems with spin-state degrees of freedom in cobalt ions, which govern the electric and magnetic properties~\cite{Tokura1998, Asai1998}.
In a trivalent cobalt ion, six electrons occupy the $3d$ orbitals.
This ion takes three distinct electronic configurations depending on the balance between Hund coupling interaction and crystalline electric-field effects: low-spin (LS) state with the $(t_{2g})^6$ configuration, intermediate-spin (IS) state with the $(t_{2g})^5(e_g)^1$ configuration, and high-spin (HS) state with the $(t_{2g})^4(e_g)^2$ configuration.
Severe competition between these states can lead to spin-state transitions or crossovers when external conditions, such as temperature or pressure, are altered.
It is well established that $\ce{LaCoO3}$ shows a spin crossover with increasing temperature~\cite{Tokura1998, Asai1998}.

The competition among the multiple spin states leads to more exotic phenomena in the presence of quantum fluctuations.
When the $t_{2g}$ and $e_g$ orbitals are regarded as valence and conduction bands, respectively, the IS state corresponds to an exciton with a triplet spin $S=1$ and the LS state is considered the vacuum of excitons~\cite{Ikeda2023, Kunes2014condensation, Kunes2014instability, Nasu2016, Kaneko2014, Kaneko2015}.
In this consideration, excitonic condensation corresponds to a quantum hybridization between the LS and IS states with long-range coherence.
This phenomenon is expected in the vicinity of the phase boundary of the LS and IS states.
As the candidate material fulfilling this condition, $\ce{Pr_{0.5}Ca_{0.5}CoO3}$ has been proposed.
This material undergoes a phase transition around 90~K without magnetic orders~\cite{Tsubouchi2002, Fujita2004, Hejtmanek2013}.
The first-principles and model calculations in Refs.~\cite{Kunes2014condensation,Kunes2014instability} suggested that the phase transition is accounted for by the spin-triplet excitonic condensation, which is also regarded as a higher-order magnetic multipole~\cite{Nasu2016,Kanekomultiploe2016,Yamaguchi2017}.

Magnetic fields offer another route to bring about competitions between the LS and IS states.
While the IS and HS states have nonzero spin moments, the LS state does not.
An applied magnetic field lowers the energy of the former but does not that of the latter, promoting a competition between the IS and LS states when the ground state without magnetic fields is the LS state.
Thus, it has been argued that an strong magnetic field applied to $\ce{LaCoO3}$ induces excitonic condensation and spin-state ordered states with different spin states arranged alternately~\cite{Altarawneh2012,Rotter2014,Ikeda2016,Ikeda2020,Ikeda2023}.
The appearance of these states is also supported by theoretical calculations~\cite{Tatsuno2016,Sotnikov2016,Kitagawa2022}.
More interestingly, a recent study suggested that ultra-high magnetic fields possibly cause a phase where both order parameters of the excitonic condensation and staggered-type spin-state alignment are nonzero simultaneously~\cite{Ikeda2023}.
This is known as an excitonic supersolid (ESS).
Nonetheless, the mechanism for stabilizing the ESS owing to applying magnetic fields remains elusive.

Moreover, orbital angular momentum and spin-orbit coupling are also expected to play a key role in discussing magnetic-field effects on electrons in $3d$ orbitals.
In cobalt oxides, both the HS and IS states retain orbital angular momentum owing to partially filled $t_{2g}$ orbitals~\cite{Kanamori1957_1, Kanamori1957_2, Tomiyasu2011}.
The spin-orbit coupling is also suggested to be crucial for interpreting the experimental results obtained from soft x-ray absorption spectroscopy and magnetic circular dichroism in $\ce{LaCoO3}$ ~\cite{Haverkort2006, Tomiyasu2017}.
Although the effects of spin-orbit coupling have been studied from the aspect of exciton condensation~\cite{Nasu2020}, comprehensive understanding including contributions of orbital moments remains elusive.

In this paper, we investigate magnetic-field effects on excitonic condensation realized in correlated electron systems with  spin-state degrees of freedom.
We introduce a half-filled two-orbital Hubbard model with a crystalline field splitting on a square lattice and examine magnetic field effects on this model comprehensively by applying the Hartree-Fock approximation.
This is a simplified model that cannot address the three spin states but can deal with competitions between LS and HS states.
In the absence of magnetic fields, we find an excitonic supersolid phase in addition to a spin-triplet excitonic phase without spin-state orderings between the LS and HS phases.
These phases with excitonic condensation are also induced by applying magnetic fields.
We also examine effects of the orbital angular momentum intrinsic to multi-orbital systems and focus on the spin-orbit coupling and contributions of orbital magnetic moments to the Zeeman terms.
We reveal that the spin-orbit coupling slightly modifies spin configurations in the spin-triplet excitonic phase but the overall structure of magnetic-field phase diagrams remains largely intact.
On the other hand, the introduction of orbital magnetic moments dramatically alters the phase boundary of the magnetic-field phase diagrams.
An applied magnetic field does not induce spin-triplet excitonic condensation for the LS phase. 
In the spin-triplet excitonic phase, the magnetic field suppresses the excitonic order parameter and changes this phase to a disordered phase connected continuously with the LS phase.
We clafity that this phenomenon originates from the orbital Zeeman effect inducing a spin-singlet excitonic order parameter, which compete with the spin-triplet excitonic phase.

This paper is organized as follows. In the next section, we introduce the two-orbital Hubbard model with a crystalline field splitting under magnetic fields.
The orbital angular momentum and spin-orbit coupling are also given in this section.
In Sec.~\ref{sec:method}, we describe the method used in the present study.
We review the Hartree-Fock approximation applied to the two-orbital Hubbard model in Sec.~\ref{sec:Hartree-Fock}.
Order parameters in this model are defined in Sec.~\ref{sec:order-parameters}.
The results are given in Sec.~\ref{sec:result}.
We show the phase diagram in the absence of magnetic fields in Sec.~\ref{sec:zero-field}.
Magnetic-field effects on the system neglecting contributions from the orbital angular momentum are examined in Sec.~\ref{sec:magnetic-field-wo-soc}.
Next, we consider the impact of the spin-orbit coupling.
After briefly discussing the phase diagram without magnetic fields in Sec.~\ref{sec:soc-zero-field}, we examine magnetic-field effects and the results are shown in Sec.~\ref{sec:soc}.
In Sec.~\ref{sec:orbital-moment}, the contribution of the orbital moment to the Zeeman term are investigated.
In Sec.~\ref{sec:discussion}, we compare the present results with previous theoretical studies and discuss the relevance to real materials.
Finally, Sec.~\ref{sec:summary} is devoted to the summary.

\section{Model}
\begin{figure}[t]
  \begin{center}
    \includegraphics[width=\columnwidth,clip]{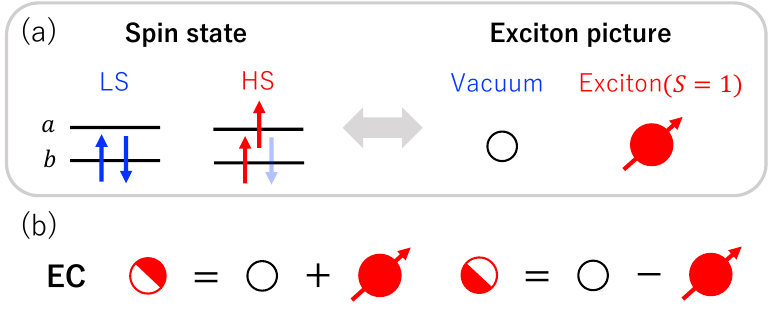}
    \caption{
    (a) Correspondence between the two spin states and their exciton pictures in the half-filled two-orbital Hubbard model~\cite{Ikeda2023}.
    The LS and HS states correspond to vacuum and an exciton state, respectively.
    (b) Schematic pictures of excitonic condensation. 
    The information on the relative phase of the superposition between the LS and HS states is presented as the direction of a semicircular shape.
    }
    \label{exciton_image}
    \end{center}
\end{figure}

In the present study, we address magnetic-field effects on excitonic condensation in correlated electron systems with orbital degrees of freedom.
In particular, we focus on the perovskite-type cobalt oxides, such as LaCoO$_3$ and Pr$_{0.5}$Ca$_{0.5}$CoO$_3$, which are the promising candidates of spin-triplet excitonic insulators.
To extract the physics based on the spin-state degrees of freedom in these cobalt oxides, the half-filled two-orbital Hubbard model with a crystalline electric-field splitting has been investigated ~\cite{Werner2007, Suzuki2009, Kanamori2011, Kanamori2012, Kunes2011}.
This model can describe two distinct spin states: the LS state with $S=0$ where two electrons occupy the lower-energy orbital, and the HS state with $S=1$ where each electron occupies a different orbital.
Furthermore, the appearance of excitonic condensation has been discussed within the two-orbital Hubbard model where the average number of electrons is two per atom~\cite{Kunes2014condensation, Kunes2014instability, Nasu2016, Kaneko2012, Kaneko2014, Kaneko2015, Niyazi2020}.
When one regards the LS state as vacuum, the HS state is interpreted as an exciton, where an electron and a hole are present in high-energy and low-energy orbitals, respectively, as shown in Fig.~\ref{exciton_image}(a).
Hybridizations between the LS and HS states [Fig.~\ref{exciton_image}(b)] are regarded as excitonic condensation.
Note that it has been pointed out that the mixing between $d_{x^2-y^2}$ in the higher-energy $e_g$ orbitals and $d_{xy}$ in the lower-energy $t_{2g}$ orbitals are crucial for the emergence of the excitonic condensation in the cobalt oxides~\cite{Kunes2014condensation,Kunes2014instability,Yamaguchi2017}.
Thus, we examine the properties of the two-orbital Hubbard model consisting of $d_{x^2-y^2}$ and $d_{xy}$ orbitals.

Although candidate materials for the spin-triplet excitonic condensation are three-dimensional compounds, we consider the two-orbital Hubbard model on a square lattice for simplicity.
This model is written as $\hat{\mathcal{H}}_{\mathrm{Hubbard}}=\hat{\mathcal{H}}_0+\hat{\mathcal{H}}_{U}$.
Here, $\hat{\mathcal{H}}_0$ represents the one-body term and given by
\begin{align}
  \hat{\mathcal{H}}_0=-\sum_{\means{ij}\eta\sigma}t_\eta \left(\hat{c}^{\dag}_{i\eta\sigma}\hat{c}_{j\eta\sigma}+\mathrm{H.c.}\right)
  +\Delta\sum_{i}\hat{n}_{ia},
  \label{eq:H0}
\end{align}
where $\means{ij}$ denotes nearest-neighbor (NN) sites, $\hat{c}_{i\eta\sigma}^\dagger$ is the creation operator of an electron occupying orbital $\eta(=a,b)$ at site $i$ with spin $\sigma(=\uparrow,\downarrow)$, and $\hat{n}_{i\eta}$ is the number operator of electrons occupying orbital $\eta$, which is given by $\hat{n}_{i\eta}=\sum_{\sigma}\hat{c}_{i\eta\sigma}^\dagger \hat{c}_{i\eta\sigma}$.
For simplicity, we only consider the electron hopping between the same orbital; $t_\eta$ is the transfer integral between orbital $\eta$.
The second term of Eq.~\eqref{eq:H0} stands for the crystalline field effect with the energy $\Delta (>0)$.
The onsite interaction term of the two-orbital Hubbard model is written as
\begin{align}
  \hat{\mathcal{H}}_{U}&=U\sum_{i\eta}\hat{n}_{i\eta\uparrow}\hat{n}_{i\eta\downarrow}+U'\sum_{i}\hat{n}_{ia}\hat{n}_{ib} \nonumber \\ 
  &+J\sum_{i\sigma\sigma'}\hat{c}^{\dag}_{ia\sigma}\hat{c}^{\dag}_{ib\sigma'}\hat{c}_{ia\sigma'}\hat{c}_{ib\sigma}
  +I\sum_{i\eta\neq \eta'}\hat{c}^{\dag}_{i\eta\uparrow}\hat{c}^{\dag}_{i\eta\downarrow}\hat{c}_{i\eta'\downarrow}\hat{c}_{i\eta'\uparrow},
  \label{eq:HU}
\end{align}
where $\hat{n}_{i\eta\sigma}=\hat{c}^{\dag}_{i\eta\sigma}\hat{c}_{i\eta\sigma}$.
The terms in Eq.~\eqref{eq:HU} with $U$, $U'$, $J$, and $I$ are the intra-orbital and inter-orbital Coulomb interactions, Hund coupling interaction, and pair-hopping interaction, respectively.

Next, we introduce the Zeeman term to examine magnetic-field effects.
This term is given by
\begin{align}
  \hat{\mathcal{H}}_{\mathrm{Zeeman}}=-\sum_{i}\bm{h}\cdot\hat{\bm{M}}_i,\label{eq:Zeeman}
\end{align}
where $\bm{h}$ is an applied magnetic field and $\hat{\bm{M}}_i$ represents the local magnetic moment at site $i$, which is written as $\hat{\bm{M}}_i=-\hat{\bm{L}}_i-2\hat{\bm{S}}_i$ with $\hat{\bm{L}}_i$ and ($\hat{\bm{S}}_i$) being orbital (spin) angular momentum.
The local spin operator $\hat{\bm{S}}_i$ is given as 
\begin{align}
    \hat{S}_i^\gamma=\sum_{\eta\sigma\sigma'}(s^\gamma)_{\sigma\sigma'}\hat{c}_{i\eta\sigma}^\dagger \hat{c}_{i\eta\sigma'},
\end{align}
where $s^\gamma=\frac12 \sigma^\gamma$ is the $2\times 2$ matrix with $\sigma^\gamma$ ($\gamma=x,y,z$) being the $\gamma$ component of the Pauli matrices.
On the other hand, the orbital angular momentum depends on the choice of the orbital basis.
In this study, we assume that $a$ and $b$ orbitals correspond to $d_{x^2-y^2}$ and $d_{xy}$ orbitals, respectively.
Since these orbitals are represented by linear combinations of the $l^z=\pm2$ states, only the $z$ component of $\hat{\bm{L}}_i$ is nonzero, which is represented as,
\begin{align}
    \hat{L}_i^z=\sum_{\eta\eta'\sigma}(l^z)_{\eta\eta'}\hat{c}_{i\eta\sigma}^\dagger \hat{c}_{i\eta\sigma}=-2i\left(\hat{c}^{\dag}_{ia\uparrow}\hat{c}_{ib\uparrow}+\hat{c}^{\dag}_{ia\downarrow}\hat{c}_{ib\downarrow}\right)
  +\mathrm{H.c.},
  \label{eq:Lz}
\end{align}
where $l^z$ is the $2\times2$ matrix given by $l^z=2\sigma^y$.
Because of the presence of orbital angular momentum, we need to consider the spin-orbit coupling.
This contribution is given by~\cite{Nasu2020}
\begin{align}
  \hat{\mathcal{H}}_\mathrm{SOC}&= \lambda\sum_{i\eta\eta'\sigma\sigma'}(l_i^z)_{\eta\eta'} (s_i^z)_{\sigma\sigma'}\hat{c}_{i\eta\sigma}^\dagger \hat{c}_{i\eta'\sigma'}\notag\\
  &=\lambda\sum_{i}\left[-i\left(\hat{c}^{\dag}_{ia\uparrow}\hat{c}_{ib\uparrow}-\hat{c}^{\dag}_{ia\downarrow}\hat{c}_{ib\downarrow}\right)+\mathrm{H.c.}\right].
  \label{eq:SOC}
\end{align}

Here, the total Hamiltonian that we should address in the presence of magnetic fields and spin-orbit coupling is $\hat{\mathcal{H}}=\hat{\mathcal{H}}_{\mathrm{Hubbard}}+\hat{\mathcal{H}}_{\mathrm{Zeeman}}+\hat{\mathcal{H}}_\mathrm{SOC}$.
We examine this Hamiltonian for the half-filled case, where average electron number is two per site.
In addition, we focus on the case with $t_a t_b >0$ corresponding to the indirect gap, which is expected in the perovskite cobalt oxides ~\cite{Yamaguchi2017, Nasu2021}.

\section{Method}
\label{sec:method}

In this section, we show the detail of the Hartree-Fock approximation that we apply to the two-orbital Hubbard model and introduce the order parameters to characterize symmetry-broken states.

\subsection{Hartree-Fock approximation}
\label{sec:Hartree-Fock}

In the Hartree-Fock approximation, we apply the following decouplings to the interaction term $\hat{\mathcal{H}}_U$:
\begin{align}
&\hat{c}^{\dag}_{i l_1}\hat{c}_{i l_2}\hat{c}^{\dag}_{i l_3}\hat{c}_{i l_4}\notag\\
&\simeq
\means{\hat{c}^{\dag}_{i l_1}\hat{c}_{i l_2}}\hat{c}^{\dag}_{i l_3}\hat{c}_{i l_4}+\hat{c}^{\dag}_{i l_1}\hat{c}_{i l_2}\means{\hat{c}^{\dag}_{i l_3}\hat{c}_{i l_4}}
-\means{\hat{c}^{\dag}_{i l_1}\hat{c}_{i l_2}}\means{\hat{c}^{\dag}_{i l_3}\hat{c}_{i l_4}} \nonumber \\
&-\means{\hat{c}^{\dag}_{i l_1}\hat{c}_{i l_4}}\hat{c}^{\dag}_{i l_3}\hat{c}_{i l_2}-\hat{c}^{\dag}_{i l_1}\hat{c}_{i l_4}\means{\hat{c}^{\dag}_{i l_3}\hat{c}_{i l_2}}
+\means{\hat{c}^{\dag}_{i l_1}\hat{c}_{i l_4}}\means{\hat{c}^{\dag}_{i l_3}\hat{c}_{i l_2}},
\label{eq:decoupling}
\end{align}
where we introduce index $l=(\eta\sigma)$ to represent orbital and spin indices together.
Here, $\means{\hat{c}^{\dag}_{i l}\hat{c}_{i l'}}$ is a local mean field.
This quantity is assumed to depend on the sublattice the site $i$ belongs to, where the number of sublattices is defined as $N_s$.
The mean field on sublattice $X$ is written as $\means{\hat{c}^{\dag}_{l}\hat{c}_{l'}}_X$.
By applying the above decoupling, the original Hamiltonian is approximated to a bilinear form of the fermionic operators as follows:
\begin{align}
  \hat{\mathcal{H}}_{\mathrm{HF}}=\sum_{\bm{k}ll'XX'} (\mathcal{M}_{\bm{k}})_{(lX),(l'X')} \hat{c}^{\dag}_{\bm{k}lX} \hat{c}_{\bm{k}l'X'} + C_0, 
\end{align}
where
$\hat{c}^\dag_{\bm{k}lX}=\sqrt{1/N_0}\sum_{i\in X}\hat{c}_{il}^\dagger e^{i\bm{k}\cdot\bm{r}_i}$ with $N_0$ and $\bm{r}_i$ being the number of unit cells and the position of site $i$, respectively, $\mathcal{M}_{\bm{k}}$ is the $4N_s\times 4N_s$ Hermitian matrix determined by the parameters of the Hamiltonian and the set of mean fields $\means{\hat{c}^{\dag}_{l}\hat{c}_{l'}}_X$, and $C_0$ is a constant.

The matrix $\mathcal{M}_{\bm{k}}$ is diagonalized by a unitary matrix $V_{\bm{k}}$ with the eigenvalues $\{\lambda_{\bm{k}1},\lambda_{\bm{k}2},\cdots\}$.
By introducing the fermionic operator $\hat{\alpha}_{\bm{k}n}=\sum_{lX}(V_{\bm{k}})_{(lX),n}^{*}\hat{c}_{\bm{k}(lX)}$, the Hartree-Fock Hamiltonian is rewritten as
\begin{align}
  \hat{\mathcal{H}}_{\mathrm{HF}}=\sum_{\bm{k}n} \lambda_{\bm{k}n} \hat{\alpha}_{\bm{k}n}^\dagger \hat{\alpha}_{\bm{k}n}+ C_0.
\end{align}
The self-consistent equation determining the mean fields is given by
\begin{align}
  \means{\hat{c}^{\dag}_l\hat{c}_{l'}}_X=\frac{1}{N_0}\sum_{\bm{k}}\means{\hat{c}^{\dag}_{\bm{k}lX}\hat{c}_{\bm{k}l'X}}
  =\frac{1}{N_0}\sum_{\bm{k}n}V^{*}_{\bm{k},(lX),n}V_{\bm{k},(l'X),n}f(\lambda_{\bm{k}n}),
\end{align}
where $f(\lambda)=1/(e^{(\lambda-\mu)/T}+1)$ is the Fermi distribution function with the chemical potential $\mu$.
We assume that the Boltzmann constant is unity.

\begin{figure}[t]
    \begin{center}
      \includegraphics[width=\columnwidth,clip]{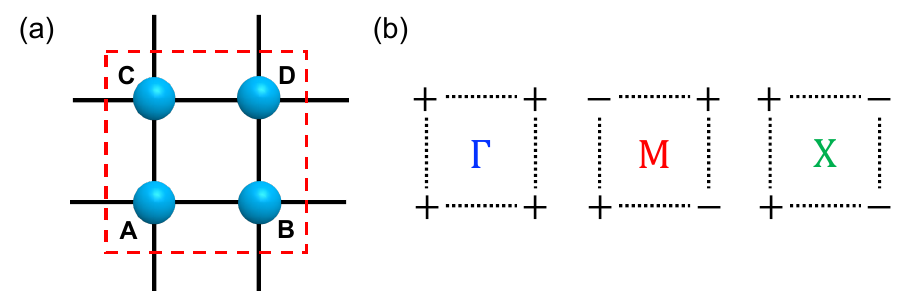}
      \caption{
      (a) Four sublattices assumed in the present Hartree-Fock calculations for a square lattice.
      The red dashed line represents a unit cell and $A$, $B$, $C$, and $D$ stand for the labels of sublattices.
      (b) Schematic pictures of three ordered structures defined in Eqs.\eqref{order_gamma}, \eqref{order_M}, and \eqref{order_X}, respectively.
      The sign $\pm$ at each site represents that of the order parameter at each sublattice.
      }
      \label{sublattice}
      \end{center}
\end{figure}

In the calculations, we impose that there are four sites in a unit cell ($N_s=4$), as shown in Fig.~\ref{sublattice}(a), and the chemical potential is determined such that the system is half-filled.
Moreover, we assume $U=U'-2J$ and $I=J$.
We focus only on the case at $T=0$ and find a stable solution with a minimal internal energy from several initial states.

\subsection{Order parameters}
\label{sec:order-parameters}

To characterize the realized state obtained by Hartree-Fock calculations, we introduce order parameters constructed from local mean fields $\means{\hat{c}^{\dag}_{l}\hat{c}_{l'}}_X$.
In the present case, there are 16 mean fields for each sublattice.
As the diagonal parts with respect to orbital $\eta$, we introduce the number of electrons and three components of the spin moment for each orbital at sublattice $X$:
\begin{align}
    n_{\eta,X} = \sum_{\sigma}\means{\hat{c}^{\dag}_{\eta\sigma}\hat{c}_{\eta\sigma}}_X,\quad
    S_{\eta,X}^\gamma = \frac12 \sum_{\sigma\sigma'}\sigma^\gamma_{\sigma\sigma'}\means{\hat{c}^{\dag}_{\eta\sigma}\hat{c}_{\eta\sigma}}_X.
\end{align}
We also define the quantities summed over the orbitals as $n_{X}=\sum_\eta n_{\eta,X}$ and $S_{X}^\gamma=\sum_\eta S_{\eta,X}^\gamma$ and the difference of the occupancy between the $a$ and $b$ orbitals as $\Delta n_{X}=n_{a,X}-n_{b,X}$.

Furthermore, we introduce the off-diagonal parts for the orbitals, which characterize excitonic condensation.
The order parameter describing the spin-triplet excitonic condensation at sublattice $X$ is defined as~\cite{Halperin1968, Kunes2015},
\begin{align}
\phi^\gamma_{t,X}=\sum_{\sigma\sigma'}\sigma^\gamma_{\sigma\sigma'}\means{\hat{c}^{\dag}_{a\sigma}\hat{c}_{b\sigma'}}_X,
  \label{order_triplet}
\end{align}
and that for the spin-singlet one is also given by
\begin{align}
  \phi_{s,X}=\sum_{\sigma}\means{\hat{c}^{\dag}_{a\sigma}\hat{c}_{b\sigma}}_X.
  \label{order_singlet}
\end{align}
While the spin-triplet excitonic condensation corresponds to the spontaneous hybridization between the LS and spin-triplet ($S=1$) HS states, the spin-singlet excitonic condensation indicates that between the LS and spin-singlet states, the latter of which is the $S=0$ state composed of two electrons singly occupying in the $a$ and $b$ orbitals. 
This implies that the Hund coupling prefers the spin-triplet excitonic condensation rather than the spin-singlet one in the the two-orbital Hubbard model.
On the other hand, it has been pointed out that the spin-singlet excitonic condensation is stabilized when the electron-lattice coupling is taken into account~\cite{Kaneko2014,Kaneko2015}.
Therefore, we mainly focus on the spin-triplet excitonic condensation in the present calculations.

The excitonic order parameters given in Eqs.~\eqref{order_triplet} and \eqref{order_singlet} take complex values.
Note that, in the subspace with the number of electrons being two at each site, the spin moment in Eq.~\eqref{order_triplet} is related to the excitonic order parameters as
\begin{align}
    S_X^\gamma\sim \mathrm{Re}\phi^{\gamma'}_{t,X}\times\mathrm{Im}\phi^{\gamma''}_{t,X},
    \label{eq:Stophi}
\end{align}
where $(\gamma,\gamma',\gamma'')=(x,y,z)$ and its cyclic permutations~\cite{Kunes2015,Nasu2020}.

As mentioned in the previous section, we perform four-sublattice calculations on the square lattice, where the sublattices are denoted as $A$, $B$, $C$, and $D$, as shown in Fig.~\ref{sublattice}(a).
To make it easier to distinguish between distinct ordered states that can be composed from these sublattice moments, we introduce three order parameters as
\begin{align}
&\mathcal{O}_{\mathrm{\Gamma}}=|\mathcal{O}_A+\mathcal{O}_B+\mathcal{O}_C+\mathcal{O}_D|, \label{order_gamma}\\
  &\mathcal{O}_{\mathrm{M}}=|\mathcal{O}_A-\mathcal{O}_B-\mathcal{O}_C+\mathcal{O}_D|, \label{order_M} \\
  &\mathcal{O}_{\mathrm{X}}=|\mathcal{O}_A-\mathcal{O}_B+\mathcal{O}_C-\mathcal{O}_D|, \label{order_X}
\end{align}
where $\mathcal{O}_X$ is an order parameter at sublattice $X$, such as $S_{\eta,X}^\gamma$ and ${\rm Re}\phi_{t,X}^\gamma$.
Note that $\mathcal{O}_{\mathrm{\Gamma}}$, $\mathcal{O}_{\mathrm{M}}$, and $\mathcal{O}_{\mathrm{X}}$ correspond to the ferro-type, antiferro-type and $(\pi,0)$-type orders for $\mathcal{O}$, respectively, as shown in Fig.~\ref{sublattice}(b).
Moreover, for simplicity, we introduce the absolute value with respect to the spin component $\gamma$ as follows:
\begin{align}
    [S]_{\rm \Gamma,M,X}=\sqrt{\sum_{\gamma=x,y,z}[S^\gamma]_{\rm \Gamma,M,X}^2}.
\end{align}
We also define $[{\rm Re}\phi_t]_{\rm \Gamma,M,X}$ and $[{\rm Im}\phi_t]_{\rm \Gamma,M,X}$ in a similar manner.

\section{Result}
\label{sec:result}

% \section{球対称な軌道での基底状態の相図}
\subsection{Phase diagram at zero magnetic field}
\label{sec:zero-field}

\begin{figure}[t]
  \begin{center}
    \includegraphics[width=\columnwidth,clip]{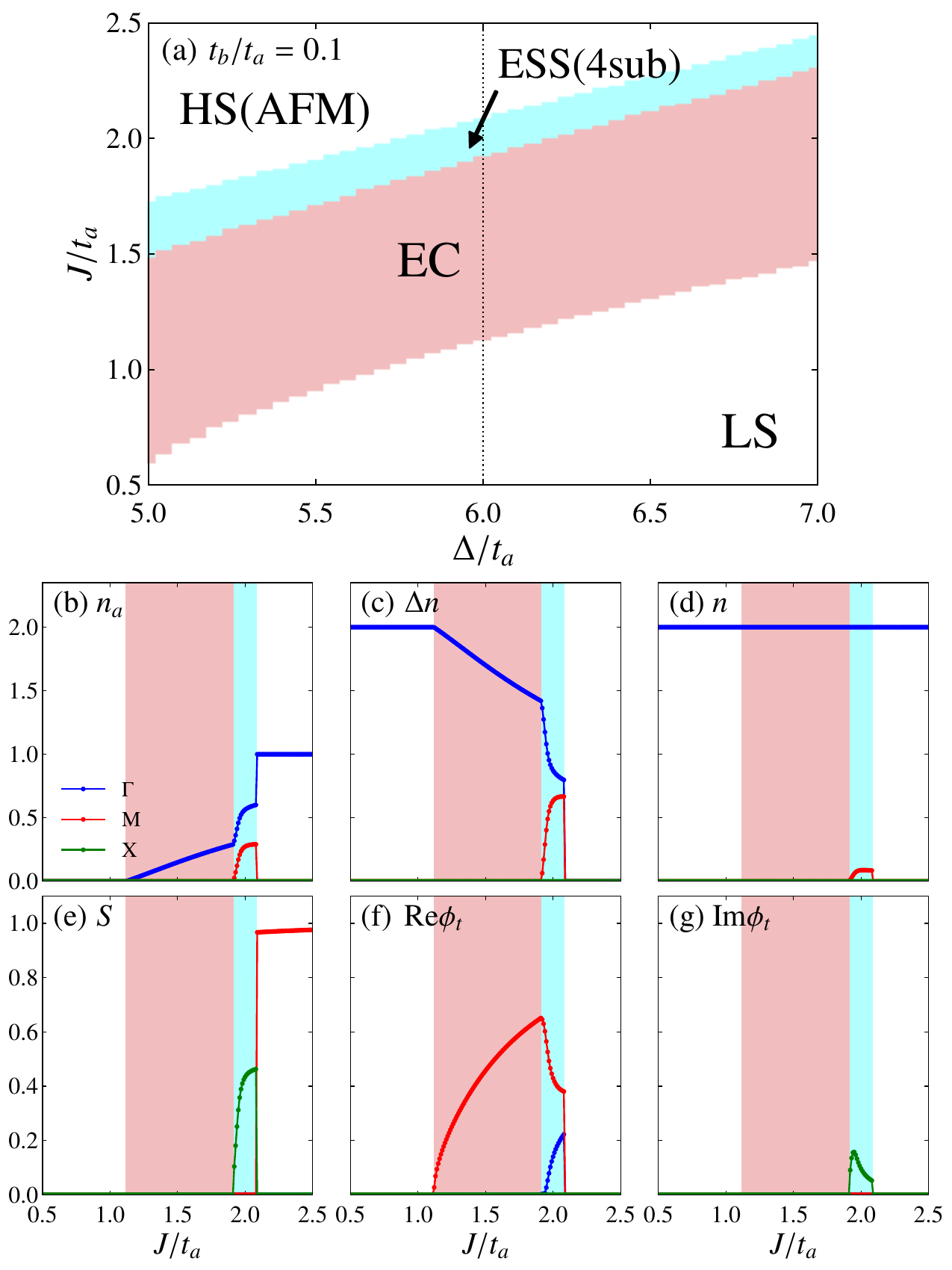}
    \caption{
    (a) Ground-state phase diagram on the plane of the Hund coupling $J$ and crystalline field $\Delta$ for $t_b/t_a=0.1$ and $U=2U'=4J$ in the absence of magnetic fields.
    Schematic pictures of four states in the phase diagram are given in Figs.~\ref{J_D_image}(a)--\ref{J_D_image}(d).
    (b)--(g) $J$ dependence of the $\Gamma$, M and X configurations of (b) the number of electrons in the $a$ orbital, (c) difference between the number of electrons in the $a$ and $b$ orbitals, (d) total number of electrons, (e) spin moment, and (f) real and (g) imaginary parts of the spin-triplet excitonic order parameters along the dotted line in (a) with $\Delta/t_a=6$.
    }
    \label{tb_0.1_J_D}
    \end{center}
\end{figure}

\begin{figure}[t]
   \begin{center}
     \includegraphics[width=1.0\columnwidth,clip]{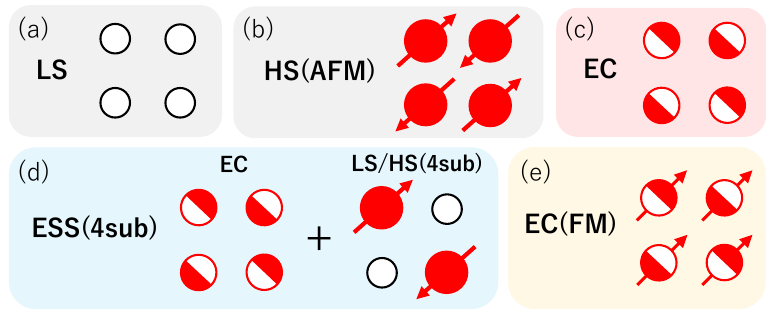}
     \caption{
     Schematic pictures of the ordered states emerging in the absence of magnetic fields for (a) the LS phase, (b) HS phase with an antiferromagnetic order, (c) spin-triplet excitonic phase, (d) four-sublattice excitonic supersolid phase, and (e) spin-triplet excitonic phase with a ferromagnetic order.
     }
     \label{J_D_image}
     \end{center}
\end{figure}

Before discussing magnetic-field effects on the two-orbital Hubbard model, we examine the properties of this model without magnetic fields.
Figure~\ref{tb_0.1_J_D}(a) shows the phase diagram on the plane of the crystalline field and Hund coupling at $t_b/t_a=0.1$ where we assume $U/J=4$ and $U'/J=2$.
In the region with the strong crystalline field $\Delta\gg J$, we find the LS phase where $n_a=0$ and $n_b=2$ are satisfied for all sites.
This is understood from the $J$ dependence of the numbers of electrons in $a$ and $b$ orbitals, as shown in Figs~\ref{tb_0.1_J_D}(b)--~\ref{tb_0.1_J_D}(d).
At $\Delta/t_a=6$, $[n_{a}]_{\Gamma}$ vanishes while $[\Delta n]_{\Gamma}=[n]_{\Gamma}=2$ below $J/t_a\simeq 1.2$, which characterizes the LS phase [Fig.~\ref{J_D_image}(a)].
On the other hand, the Hund coupling is strong enough, we find the HS phase with the antiferromagnetic (AFM) order [Fig.~\ref{J_D_image}(b)], where $n_a=1$ and $n_b=1$ are satisfied for all sites~\cite{Werner2007,Kanamori2011, Kanamori2012, Nasu2016, Hoshino2016}, which are confirmed from the results in Figs~\ref{tb_0.1_J_D}(b)--~\ref{tb_0.1_J_D}(d).
The AFM order in the HS phase is determined from the fact that $[S]_{\rm M}$ takes a nonzero value in this region [Fig~\ref{tb_0.1_J_D}(e)].

In between the LS and HS phases, we find two phases with excitonic condensation due to the competition between the crystalline-field effect and Hund coupling interaction.
One is the spin-triplet excitonic phase characterized by the nonzero ${\rm Re}\phi_t$, where the LS and HS states are spontaneously hybridized as illustrated in Fig.~\ref{exciton_image}(b).
The $J$ dependence of this quantity at $\Delta/t_a=6$ is presented in Fig.~\ref{tb_0.1_J_D}(f).
In the phase termed EC in Fig.~\ref{tb_0.1_J_D}(a), $[{\rm Re}\phi_t]_{\rm M}$ takes a nonzero value, indicating an antiferro-type excitonic order, as shown in Fig.~\ref{J_D_image}(c).
The antiferro-type order is attributed to the indirect-gap band structure with $t_a t_b>0$ ~\cite{Kunes2014instability, Kunes2015, Kaneko2012, Kaneko2014, Kaneko2015, Yamaguchi2017, Li2020}. When the direct gap with $t_a t_b<0$ is assumed, a ferro-type excitonic order is expected as discussed in the previous studies ~\cite{Kunes2015, Kunes2014instability, Nasu2016, Yamamoto2020}. Note that ${\rm Im}\phi_t$ is zero, as shown in Fig.~\ref{tb_0.1_J_D}(g), in the EC phase.
This originates from the pair-hopping interaction, which fixes the phase of $\phi_t$ to zero~\cite{Kunes2014instability, Kaneko2015,Nasu2016}.

In addition to the EC phase, we find another excitonic phase with nonzero $\phi_t$ [see Fig.~\ref{tb_0.1_J_D}(f)] near the HS phase.
This excitonic phase is distinguished from the EC phase by the presence of nonzero $[\Delta n]_{\rm M}$ [see Fig.~\ref{tb_0.1_J_D}(c)] indicating the staggered-type spin-state order corresponding to a superlattice structure formed by excitons.
Since this excitonic phase is characterized by condensation and superlattice formation of excitons emergent simultaneously, it is regarded as a supersolid of excitons, termed ESS in Fig.~\ref{tb_0.1_J_D}(a).
In addition to nonzero $[\Delta n]_{\rm M}$ and $[{\rm Re} \phi_t]_{\rm M}$, we find the $(\pi,0)$-type spin order characterized by the appearance of $[S]_{\rm X}$, as shown in Fig.~\ref{tb_0.1_J_D}(e).
The $(\pi,0)$-type spin order comes from the antiferromagnetic correlation between HS states, which aligned alternatively [see Fig.~\ref{J_D_image}(d)].
Because of the coexistence of the antiferro-type excitonic and $(\pi,0)$-type spin orders, this phase is characterized as a four-sublattice order and denoted as ESS(4sub) in Fig.~\ref{tb_0.1_J_D}(a).
From Eq.~\eqref{eq:Stophi}, the $(\pi,0)$-type spin order accompanied by the antiferro-type excitonic order with nonzero $[{\rm Re} \phi_t]_{\rm M}$ results in nonzero $[{\rm Im} \phi_t]_{\rm X}$, as shown in Fig.~\ref{tb_0.1_J_D}(g).
Since the magnitude of the quantum hybridization between the LS and HS states for HS dominant sites is different from that for LS dominant sites, the uniform component of the excitonic-order parameter, $[{\rm Re} \phi_t]_{\Gamma}$, becomes nonzero in the ESS phase.
Furthermore, the superlattice formation of excitons results in a charge disproportionation, which is characterized by $[n]_{\rm M}$ as shown in Fig.~\ref{tb_0.1_J_D}(d).
Note that the phase transition between HS and ESS phases is of first order but those betwen the LS and EC phases and the EC and ESS phases appear to be of second order.
The origin of the first-order transition is ascribed to the different spin-order patterns in the ESS and HS phases.

The emergence of a phase similar to ESS has been suggested by dynamical mean-field theory (DMFT) at finite temperatures in the two orbital Hubbard model where the spin exchange in the Hund coupling and pair-hopping terms are neglected for simplicity~\cite{Niyazi2020}.
The previous study have revealed that the ESS appears in a very narrow region at the boundary between the EC and spin-state ordered phase, which is different from the present result; we do not find the spin-state ordered phase without excitonic condensation.
We believe that the difference is due to the simplification of the previous study because the pair-hopping interaction fixing the phase of $\phi_t$ stabilizes excitonic condensation.
Indeed, we have confirmed that neglecting the pair-hopping term stabilizes the spin-state ordered order without excitonic condensation.
This result is consistent with the previous study using a variational cluster approximation, which clarified that the pair-hopping term stabilizes the spin-triplet excitonic condensation~\cite{Kaneko2014, Kaneko2015}.

\begin{figure}[t]
  \begin{center}
    \includegraphics[width=\columnwidth,clip]{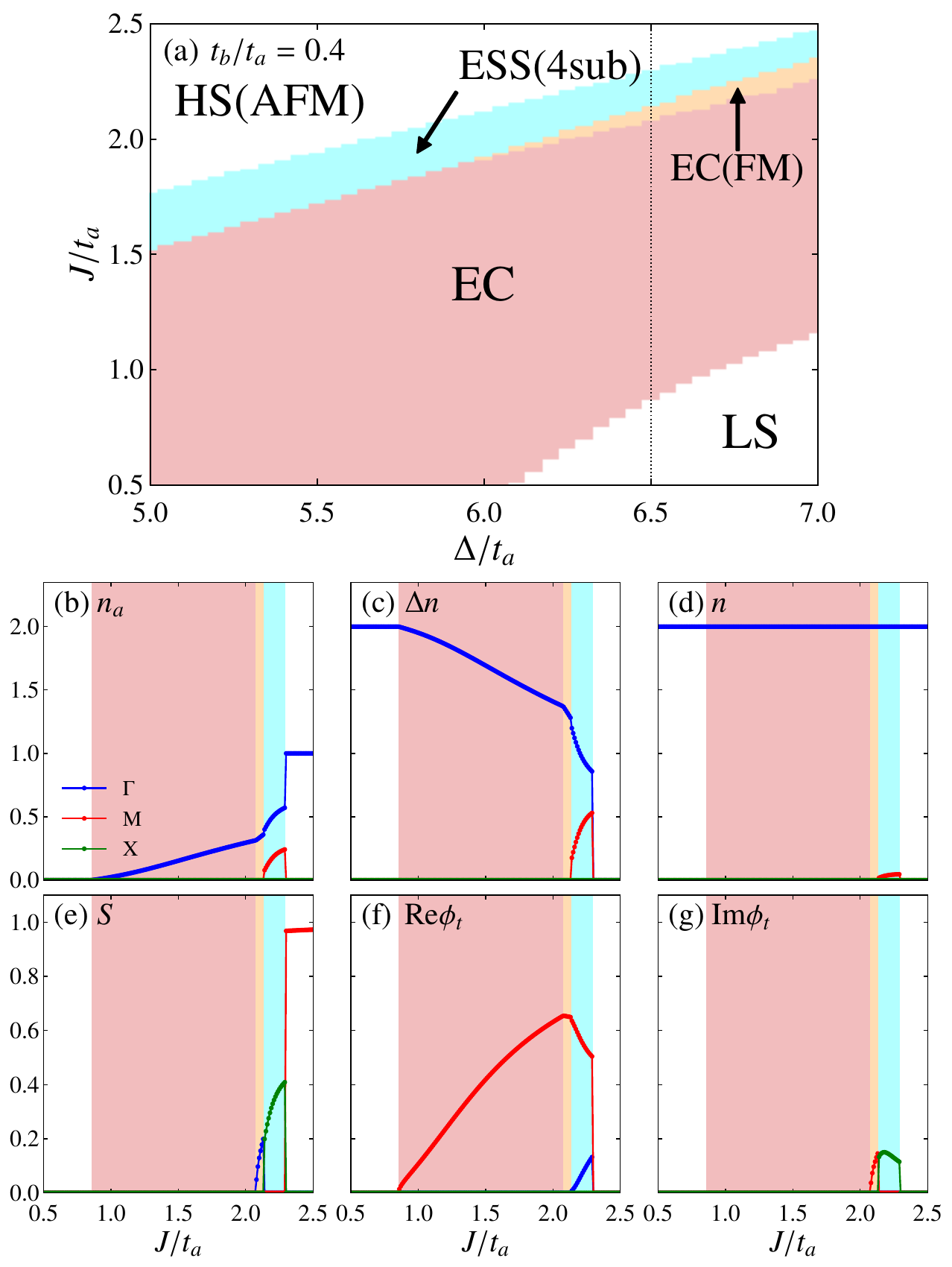}
    \caption{
         Similar plots to Fig.~\ref{tb_0.1_J_D} for the case with $t_b/t_a=0.4$.
         The $J$ dependence of the order parameters in (b)--(g) is calculated along the dotted line in (a) with $\Delta/t_a=6.5$.
    }
    \label{tb_0.4_J_D}
    \end{center}
\end{figure}

Next, we examine the effect of the ratio between the bandwidths of the two orbitals.
Figure~\ref{tb_0.4_J_D}(a) shows the phase diagram at $t_b/t_a=0.4$ and other parameters are the same as that in Fig.~\ref{tb_0.1_J_D}.
The region of the EC phase is larger than that at $t_b/t_a=0.1$ presented in Fig.~\ref{tb_0.1_J_D}(a).
This tendency can be understood from the fact that the itinerancy of excitons is proportional to the product of $t_a$ and $t_b$ ~\cite{Kunes2014instability, Nasu2016}.
Moreover, we find a ferromagnetic phase accompanied by excitonic condensation between the EC and ESS phases, which is schematically depicted in Fig.~\ref{J_D_image}(e).
Figures~\ref{tb_0.4_J_D}(b)--\ref{tb_0.4_J_D}(g) present the $J$ dependence of the order parameters along the dotted line in Fig.~\ref{tb_0.4_J_D}(a) at $\Delta/t_a=6.5$.
Around $J/t_a=2.2$, there is a region where both $[S]_{\Gamma}$ and $[{\rm Re}\phi_t]_{\rm M}$ are nonzero simultaneously, which corresponds to the ferromagnetic EC phase [Figs.~\ref{tb_0.4_J_D}(e) and \ref{tb_0.4_J_D}(f)].
In this phase, $[{\rm Im}\phi_t]_{\rm M}$ is also nonzero because of the relation given in Eq.~\eqref{eq:Stophi} [Fig.~\ref{tb_0.4_J_D}(g)].
The previous DMFT studies for the two-orbital model without the pair-hopping interaction and spin exchange part of the Hund coupling interaction suggested the presence of a ferromagnetic instability in the excitonic phase but did not find an excitonic phase with a ferromagnetic order under the half-filled condition~\cite{Niyazi2020,Geffroy2019}.
The difference from the previous study may be due to the artifact of the Hartree-Fock approximation in the present calculations, which could overestimate effects of the Hund coupling interaction leading to ferromagnetism.
It is difficult to settle the presence of a ferromagnetic excitonic phase in the half-filled Hubbard model, and this issue remains a future work.

\subsection{Magnetic-field effect without orbital angular momentum}
\label{sec:magnetic-field-wo-soc}

\begin{figure}[t]
  \begin{center}
    \includegraphics[width=\columnwidth,clip]{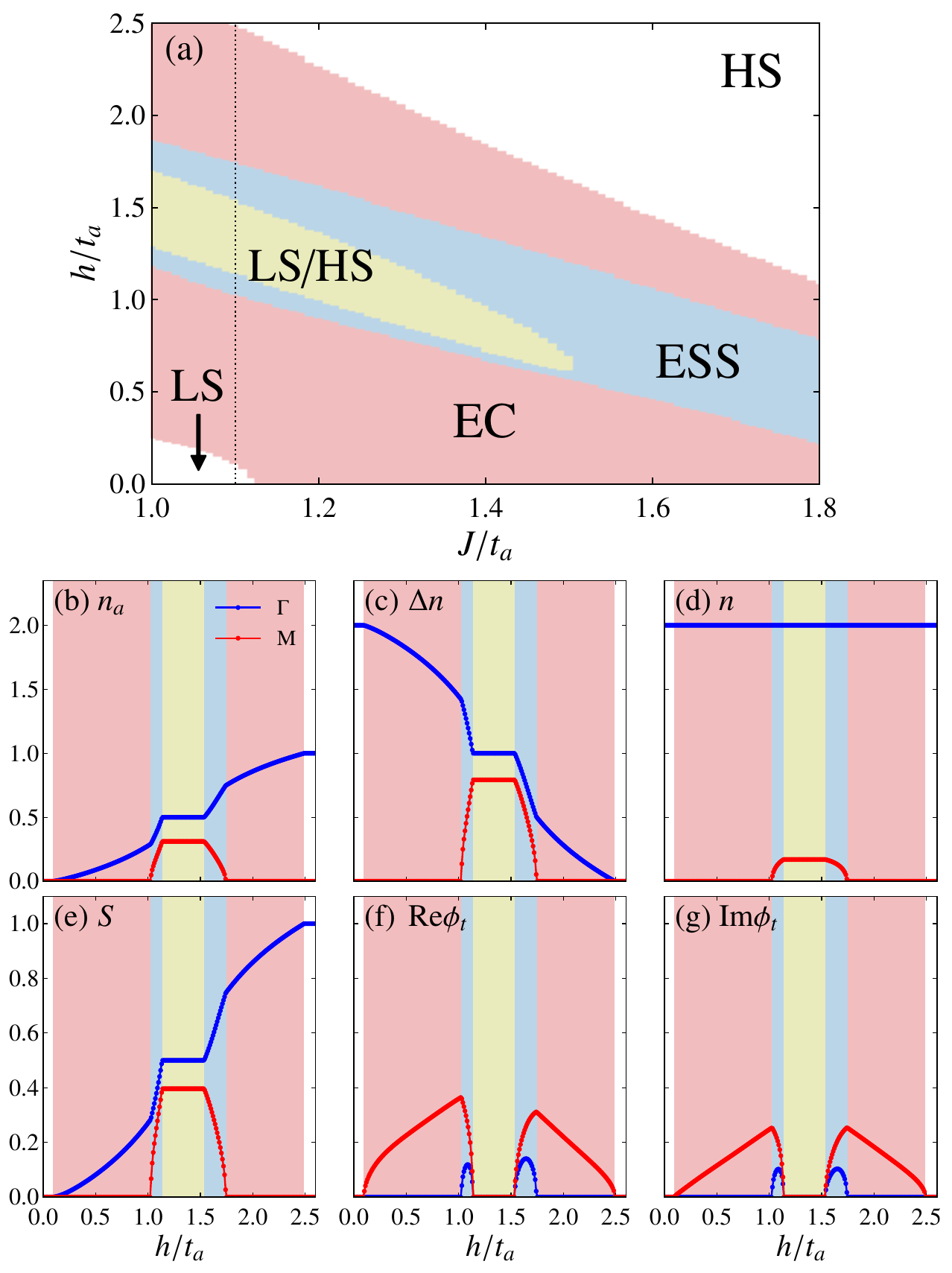}
    \caption{
    (a) Ground-state phase diagram on the plane of the Hund coupling $J$ and magnetic field $h$ for $t_b/t_a=0.1$, $\Delta/t_a=6$, and $U=2U'=4J$.
    Schematic pictures of four states in the phase diagram are given in Fig.~\ref{h_image}.
    (b)--(g) Magnetic-field dependence of the $\Gamma$ and M configurations of (b) the number of electrons in the $a$ orbital, (c) difference between the number of electrons in the $a$ and $b$ orbitals, (d) total number of electrons, (e) spin moment, and (f) real and (g) imaginary parts of the spin-triplet excitonic order parameters along the dotted line in (a) with $J/t_a=1.1$.
    }
    \label{J_h}
    \end{center}
\end{figure}

\begin{figure}[t]
  \begin{center}
    \includegraphics[width=\columnwidth,clip]{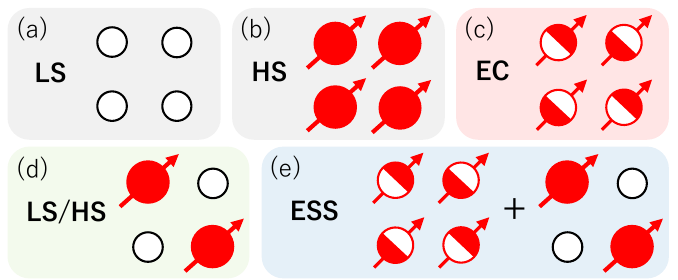}
    \caption{
     Schematic pictures of the ordered states emerging in the presence of magnetic fields for (a) the LS phase, (b) spin-polarized HS phase, (c) spin-triplet excitonic phase, (d) staggered-type LS-HS ordered phase, and (e) two-sublattice excitonic supersolid phase.
    }
    \label{h_image}
    \end{center}
\end{figure}

Here, we present the results under magnetic fields.
In this section, we omit the effects originating from the orbital angular momentum given in Eq.~\eqref{eq:Lz}, which corresponds to neglecting the contribution of the orbital magnetic moment to the Zeeman term in Eq.~\eqref{eq:Zeeman} and the spin-orbit coupling in Eq.~\eqref{eq:SOC}.
This assumption results in the absence of the angle dependence of an applied magnetic field and parameterize the magnetic-field effect using its magnitude $h$.
Figure~\ref{J_h}(a) shows the phase diagram on the plane of the Hund coupling $J$ and field magnitude $h$, where $t_b/t_a=0.1$ and $\Delta/t_a=6$.
In the absence of the magnetic field, the LS, EC, ESS, and HS phases appear by changing $J$, as illustrated in Fig.~\ref{tb_0.1_J_D}.
Note that the EC state possesses no local spin moments, but ESS and HS phases exhibit magnetic orders with no total magnetization.

Applying the magnetic field to the LS phase brings about the phase transition into an EC phase [see Fig.~\ref{J_h}(b)--(g)].
In the field-induced EC phase, a nonzero total spin moment emerges [Fig.~\ref{J_h}(e)] in addition to $[{\rm Re}\phi_t]_{\rm M}\ne 0$, as shown in Fig.~\ref{J_h}(f).
The schematic picture of this state is depicted in Fig.~\ref{h_image}(c).
As the result of the spin moment induced by the magnetic field, $\mathrm{Im}\phi_t$ is also nonzero, and three vectors $\bm{h}$ (parallel to the local spin moment), $\mathrm{Re}\vec{\phi}_t$ and $\mathrm{Im}\vec{\phi}_t$ are orthogonal to each other at each site owing to Eq.~\eqref{eq:Stophi}.
Further increase of the magnetic field results in the state with nonzero $[\Delta n]_{\rm M}$, corresponding to the staggered alignment of the LS and HS states [see Fig.~\ref{J_h}(c)].
The continuous phase transition occurs from the EC to ESS where both $[{\rm Re}\phi_t]_{\rm M}$ and $[\Delta n]_{\rm M}$ are nonzero.
Unlike the four-sublattice ESS phase appearing in the absence of magnetic fields, this field-induced ESS phase is described as a two-sublattice order where spin moments in HS sites are in the same direction, as shown in Fig.~\ref{h_image}(e).

With increasing the magnetic field, the excitonic order parameters vanish, and the phase characterized by nonzero $[\Delta n]_{\rm M}$ appears, indicated by the yellow region in Fig.~\ref{J_h}.
This phase is regarded as an LS-HS superlattice corresponding to a staggered order of excitons [Fig.~\ref{h_image}(d)], which is denoted by LS/HS in Fig.~\ref{J_h}(a).
We find that the staggered-type charge disproportionation emerges in the LS/HS phase in addition to the ESS phase [Fig.~\ref{J_h}(d)].
The detailed analysis clarifies that the number of electrons in LS sites is more than that in HS sites.
This is understood from the kinetic-energy gain of spin-polarized electrons in the $a$ orbital while the system remains insulating.
Thus, an applied magnetic field causes charge disproportionations in the phases with an LS-HS order.
The disproportionation in the field-induced ESS phase is also understood in a similar manner.

Further increase of the magnetic field results in reentry into the ESS and EC phases successively, and finally, the system becomes the fully spin-polarized HS phase, as depicted in Fig.~\ref{h_image}(b).
The magnetization curve is presented in Fig.~\ref{J_h}(e).
It changes continuously as a function of an applied magnetic field, and we find a plateau at half of the maximum value in the LS/HS phase.
When the magnetization takes its maximum, a fully spin-polarized HS phase appears, as shown in Fig.~\ref{h_image}(b).

\subsection{Effect of spin-orbit coupling at zero magnetic field}
\label{sec:soc-zero-field}

\begin{figure}[t]
  \begin{center}
    \includegraphics[width=\columnwidth,clip]{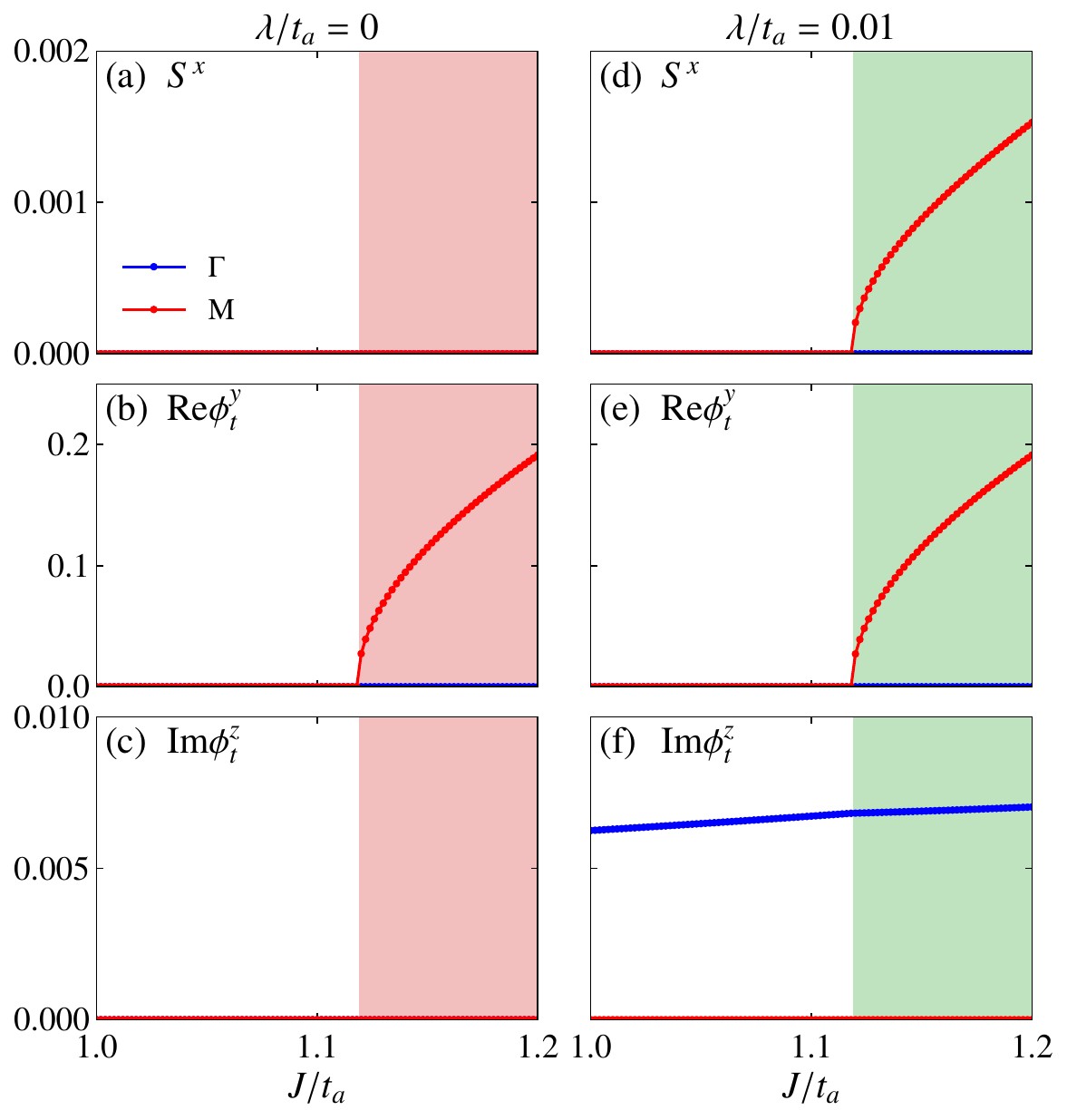}
    \caption{
    (a)--(c) $J$ dependence of (a) $S^x$, (b) $\mathrm{Re}\phi^y_t$, and (c) $\mathrm{Im}\phi^z_t$ for the ferro-type ($\Gamma$) and antiferro-type (M) configurations without the spin-orbit coupling.
    (d)--(f) Similar plots for the system with $\lambda/t_a=0.01$.
    The other parameters are set to $t_b/t_a=0.1$, $\Delta/t_a=6$, and $U=2U'=4J$.
    }
    \label{l_J}
    \end{center}
\end{figure}

In this section, we examine the properties of systems with the spin-orbit coupling and consider the indirect-gap system with $t_a t_b>0$, which is different from previous calculations in Ref.~\cite{Nasu2020} where $t_a t_b<0$ is assumed.
The systems with direct and indirect gaps are connected via transformation with $c_{ib\sigma}\rightarrow(-1)^ic_{ib\sigma}$ in the absence of the spin-orbit coupling ~\cite{Kunes2015}.
Nonzero spin-orbit coupling violates this connection, implying distinct phase diagrams for these two cases.
It is also worth noting that $\hat{\cal H}_{\rm SOC}$ is expected to induce $[\mathrm{Im}\phi^z_t]_{\Gamma}$ from Eq.~\eqref{eq:SOC}.
A change in the sign of $\lambda$ affects only the induced $[\mathrm{Im}\phi^z_t]_{\Gamma}$ and does not alter the overall phase diagram.
Figure~\ref{l_J} shows the $J$ dependence of relevant order parameters in the absence and presence of the spin-orbit coupling.
In this absence, the EC phase characterizing nonzero $[\mathrm{Re}\phi^z_t]_{\rm M}$ does not possess any magnetic orders [see Fig.~\ref{l_J}(a)--\ref{l_J}(c)].
When the spin-orbit coupling is introduced, $[\mathrm{Im}\phi^z_t]_{\Gamma}$ is always nonzero even in the LS phases [see Fig.~\ref{l_J}(e) and \ref{l_J}(f)].
We find that the EC phase is accompanied by an antiferromagnetic order with tiny spin moments, as shown in Fig.~\ref{l_J}(d).
The magnetic order along the $S^x$ direction arises from the coexistence of the two nonzero parameters $[\mathrm{Re}\phi^y_t]_{\rm M}$ and $[\mathrm{Im}\phi^z_t]_{\Gamma}$ [see Eq.~\eqref{eq:Stophi}].
Note that there is a rotational symmetry around the $S^z$ axis, and therefore, the EC phase with $[\mathrm{Re}\phi^x_t]_{\rm M}$ results in the antiferromagnetic order along the $S^y$ direction.
On the other hand, if the EC phase appears with only the $z$ component of $\mathrm{Re}\phi_t$, the spin-orbit coupling does not induce local spin moments, which is understood from Eq.~\eqref{eq:Stophi}.
Since the emergence of spin moments gives the energy gain of antiferromagnetic interactions between neighboring HS sites, the excitonic order parameter of the EC phase prefers lying on the $\mathrm{Re}\phi_t^x$-$\mathrm{Re}\phi_t^y$ plane and is associated with local magnetic moments on the $S^x$-$S^y$ plane.

\subsection{Magnetic-field effect with spin-orbit coupling}
\label{sec:soc}

\begin{figure}[t]
  \begin{center}
    \includegraphics[width=\columnwidth,clip]{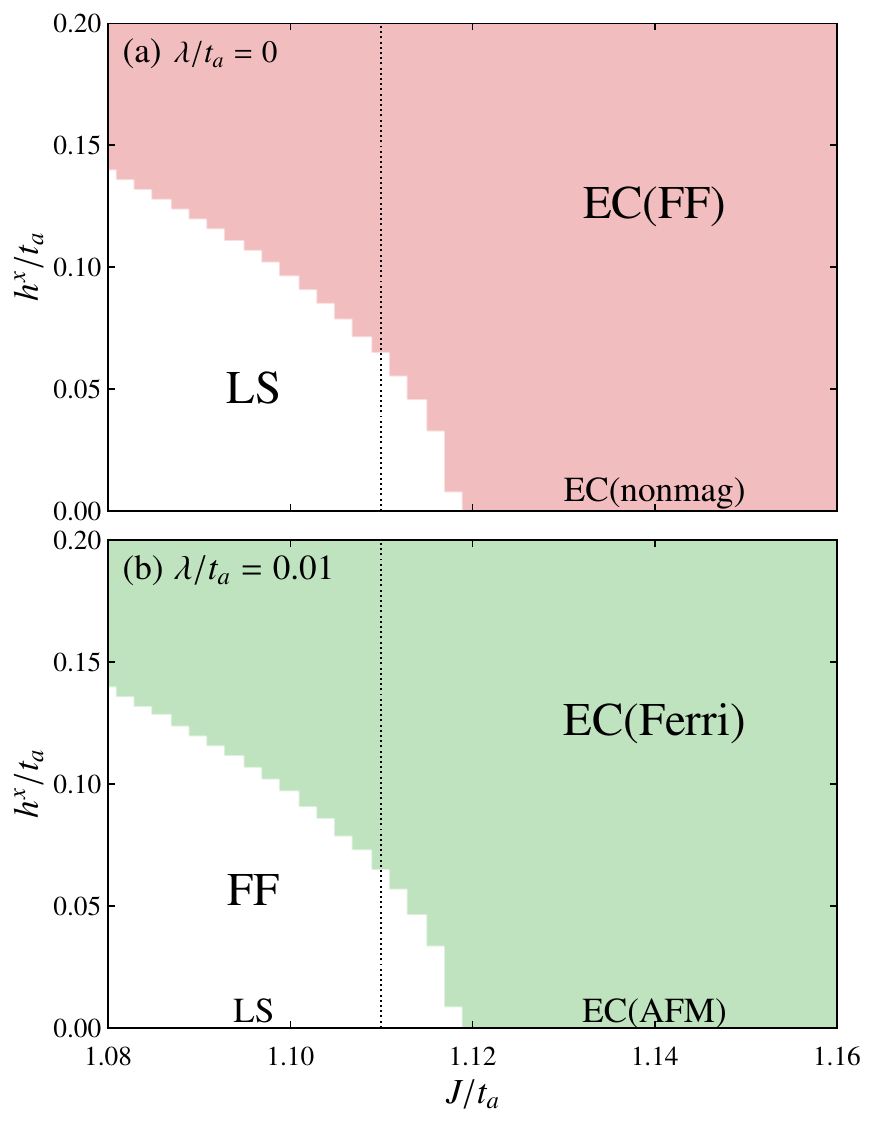}
    \caption{
    Ground-state phase diagrams on the plane of the Hund coupling $J$ and magnetic field $h^x$ for (a) $\lambda/t_a=0$ and (b) $\lambda/t_a=0.01$.
    The other parameters are set to $t_b/t_a=0.1$, $\Delta/t_a=6$, and $U=2U'=4J$.
    }
    \label{J_hx_SOC}
    \end{center}
\end{figure}

\begin{figure}[t]
  \begin{center}
    \includegraphics[width=\columnwidth,clip]{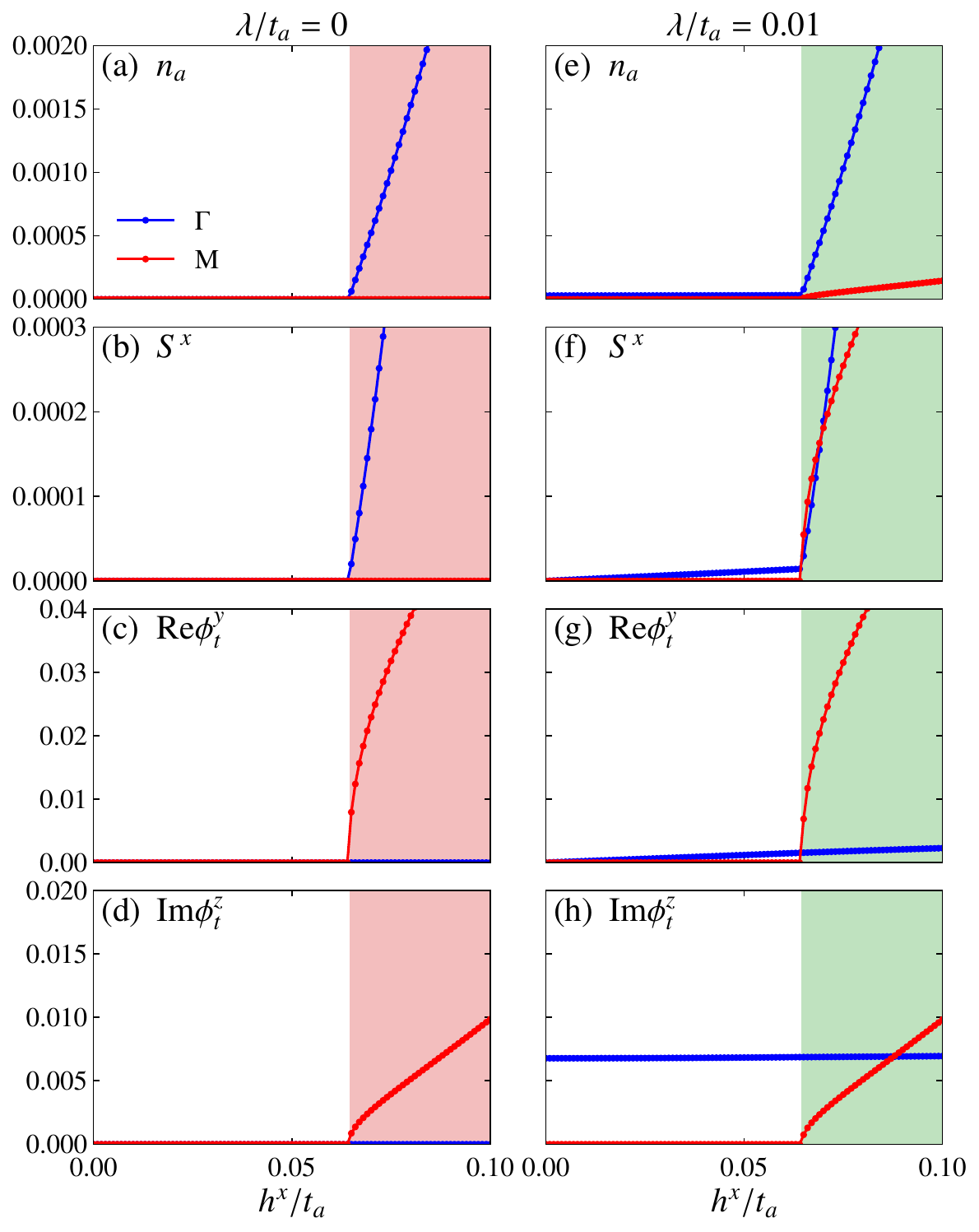}
    \caption{
    (a)--(d) Magnetic-field dependence of (a) $n_a$, (b) $S^x$, (c) $\mathrm{Re}\phi^y_t$, and (d) $\mathrm{Im}\phi^z_t$ for the ferro-type ($\Gamma$) and antiferro-type (M) configurations without the spin-orbit coupling.
    The other parameters are set to $t_b/t_a=0.1$, $\Delta/t_a=6$, and $U=2U'=4J$.
    (e)--(h) Similar plots for $\lambda/t_a=0.01$.
    The calculations are made along the dotted lines in Fig.~\ref{J_hx_SOC} with $J/t_a=1.11$.
    }
    \label{l_hx}
    \end{center}
\end{figure}

As discussed in the previous section, the inplane spin anisotropy on the $S^x$-$S^y$ plane is present in the EC phase in the absence of magnetic fields.
This suggests that magnetic fields applying along the $h^x$ or $h^y$ directions cause different results from those in fields along the $h^z$ direction.
Here, we focus on the effects of the magnetic field along the $h^x$ direction.
Figures~\ref{J_hx_SOC}(a) and \ref{J_hx_SOC}(b) show the phase diagrams in the absence and presence of the spin-orbit coupling, respectively.
The phase boundary between the LS and the EC phases remains largely unaffected by the introduction of the spin-orbit coupling.
However, their magnetic properties are critically different from each other.
Figure~\ref{l_hx} shows the $h^x$ dependence of order parameters obtained along the dotted lines in Fig.~\ref{J_hx_SOC}.
% A critical magnetic field demarcates the boundary between the two phases.
In the scenario with $\lambda=0$, the lower phase is composed only of the pure LS states as indicated by $[n_a]_{\Gamma}=0$ in Fig.~\ref{l_hx}(a).
Conversely, a nonzero $\lambda$ leads to a hybridization between the LS and HS states, and $[\mathrm{Im}\phi^{z}_t]_{\Gamma}$ and $[n_a]_{\Gamma}$ become nonzero, as seen in Figs.~\ref{l_hx}(e) and \ref{l_hx}(h).
Due to the spin-state hybridization by the spin-orbit coupling, the applied magnetic field induces a spin polarization [Fig.~\ref{l_hx}(f)], which is distinctly different from the case with $\lambda=0$ [Fig.~\ref{l_hx}(b)].
The spin polarization leads to ferro-type $\mathrm{Re}\phi^{y}_t$ as shown in Fig.~\ref{l_hx}(g) [see Eq.~\eqref{eq:Stophi}].
Therefore, the low-field phase in Fig.~\ref{J_hx_SOC}(b) is regarded as the forced ferromagnetic (FF) phase, and the magnetic susceptibility is nonzero, unlike the LS phase without spin-orbit coupling.
This result is consistent with the previous study where $t_at_b<0$ is assumed ~\cite{Nasu2020}.

Next, we focus on the higher-field EC phase, which is defined by a nonzero $[\mathrm{Re}\phi^{y}_t]_{\rm M}$ independent of the existence of the spin-orbit coupling [Figs.~\ref{l_hx}(c) and \ref{l_hx}(g)].
In the presence of the spin-orbit coupling, the coexistence of $[\mathrm{Im}\phi^{z}_t]_{\Gamma}$ and $[\mathrm{Re}\phi^{y}_t]_{\rm M}$ yields the antiferromagnetic order of the $S^x$ component despite the magnetic field applied along the $h^x$ direction [see Figs.~\ref{l_hx}(f)--(h)].
Thus, the applied magnetic field induces a ferrimagnetic state aligned with the field direction, rather than spin canting.
Note that the field-induced ferrimagnetic EC phase also possesses a staggered-type spin-state order in the presence of the spin-orbit coupling, as shown in Fig.~\ref{l_hx}(e).

We comment on the results under magnetic fields along the $h^z$ direction.
Similar to the results with the field along the $h^x$ direction shown in Fig.~\ref{J_hx_SOC}(b), the phase boundary on the plane of $J$ and $h^z$ remains largely intact even in the presence of the spin-orbit coupling.
Nevertheless, the spin alignments in the two phases differ from those in magnetic fields along the $h^x$ direction.
The in-plane anisotropy, which arises from the spin-orbit coupling, prevents the emergence of spin moments due to the magnetic field in the lower-field phase.
On the other hand, in the higher-field EC phase, the magnetic field along the $h^z$ direction induces the spin moment while this EC phase is associated with antiferromagnetic order on the $S^x$-$S^y$ plane.
Thus, a spin canting is observed in the EC phase.

\subsection{Effect of orbital magnetic moment}
\label{sec:orbital-moment}

\begin{figure}[t]
  \begin{center}
    \includegraphics[width=\columnwidth,clip]{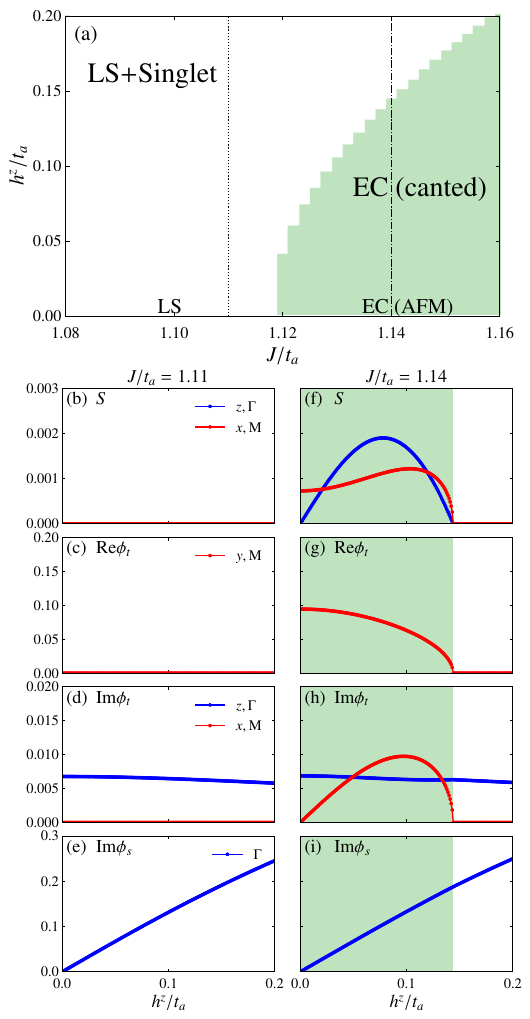}
    \caption{
    (a) Ground-state phase diagram on the plane of the Hund coupling $J$ and magnetic field $h^z$ for $\lambda/t_a=0.01$, where the contribution of the orbital angular momentum to the magnetic moment is taken into account. 
    The other parameters are set to $t_b/t_a=0.1$, $\Delta/t_a=6$, and $U=2U'=4J$.
    (b)--(e) Magnetic-field dependence of (b) $[S^z]_\Gamma$ and $[S^x]_{\rm M}$, (c) $[\mathrm{Re}\phi_t^y]_{\rm M}$, (d) $[\mathrm{Im}\phi_t^z]_\Gamma$ and $[\mathrm{Im}\phi_t^x]_{\rm M}$, and (e) $[\mathrm{Im}\phi_s]_\Gamma$ for $J/t_a=1.11$ along the dotted line in (a).
    (f)--(i) Similar plots to (b)--(e) for $J/t_a=1.14$ along the dashed-dotted line in (a).
    }
    \label{l_hz}
    \end{center}
\end{figure}

Finally, we examine the effect of the orbital angular momentum in the Zeeman term given in Eq.~\eqref{eq:Zeeman} in addition to the spin-orbit coupling.
Since only the $z$ component is nonzero, we here consider the case with magnetic fields applied along the $h^z$ direction.
We have confirmed that the results obtained for the other field directions are identical to those in Fig.~\ref{J_hx_SOC}.
The Zeeman effect on the orbital moment induces the spin-singlet excitonic order parameter given in Eq.~\eqref{order_singlet} because $\means{\hat{L}^z}\propto \mathrm{Im}\phi_s$.
In other words, this effect mixes the LS state with the spin-singlet state, although the spin-orbit coupling results in the hybridization between the LS state and the HS spin-triplet states.
Thus, the orbital contribution to the Zeeman effect competes with the appearance of the spin-triplet excitonic condensation.
Figure~\ref{l_hz}(a) presents the magnetic-field phase diagram when the spin-orbit coupling and the effect of the orbital angular momentum are taken into account.
In the absence of magnetic fields, the LS and EC phases with antiferromagnetic order appear depending on the Hund coupling, as in Fig.~\ref{J_hx_SOC}(b).
While the magnetic field changes the LS phase to the excitonic phase in the system without the orbital angular momentum, such a field-induced transition does not occur for the LS phase, as shown in Fig.~\ref{l_hz}(a).
The spin moment is not induced in the LS phase, as shown in Fig.~\ref{l_hz}(b).
Instead, the orbital magnetic moment proportional to $\mathrm{Im}\phi_s$ appears as shown in Fig.~\ref{l_hz}(e).
This is attributed to the fact that the $d_{x^2-y^2}$ and $d_{xy}$ orbitals possess a large orbital moment with $l^z=\pm 2$, suggesting that the doubly occupied $l^z=- 2$ state with $M^z=4$ is preferred by strong magnetic fields along $h^z$ direction rather than the $S=1$ spin triplet state with $M^z=2$.

Next, we discuss magnetic-field effects on the EC phase.
Figures~\ref{l_hz}(f)--\ref{l_hz}(i) show the $h^z$ dependence of the order parameters.
In the absence of the magnetic field, the EC phase is associated with the antiferromagnetic order along the $S^x$ direction, as shown in Fig.~\ref{l_hz}(f).
An applied magnetic field induces the spin moment along the $S^z$ direction, and a canted spin structure is realized, similar to the case without the orbital angular momentum.
We also find that the EC order parameter is suppressed with increasing $h^z$ and vanishes above a critical field, as shown in Fig.~\ref{l_hz}(g).
Above the critical field, the spin moment also disappears, and the higher-field phase is continuously connected with the LS phase.
Although the spin moment vanishes, $\mathrm{Im}\phi_s$ corresponding to the orbital moment monotonically increases with increasing the magnetic field [Fig.~\ref{l_hz}(i)].
This is ascribed to the dominant contribution of the orbital moment to the magnetization.

\section{Discussion}
\label{sec:discussion}

Here, we compare the current results with the previous studies and discuss their relevance to real materials such as perovskite cobalt oxides.
In the previous study, the impact of magnetic fields on the two-orbital Hubbard model has been examined using DMFT, where the pair hopping interaction and spin-exchange terms were neglected, as mentioned in Ref.~\cite{Sotnikov2016}.
This study has revealed that an excitonic phase is induced by applying a magnetic field to the LS phase, consistent with our findings where orbital magnetic moments are omitted. 
Meanwhile, the field-induced ESS phase was not observed in the earlier study.
Given the detailed DMFT calculations conducted without magnetic fields, which indicated the presence of the ESS phase as discussed in Sec.~\ref{sec:zero-field}, we expect that this phase would emerge in a region of higher magnetic fields.

The effects of magnetic fields on the two-orbital Hubbard model in the strong correlation limit have also been examined so far~\cite{Tatsuno2016}.
In this approach, a localized quantum model is introduced using a second-order perturbation theory, where the electron hopping between neighboring sites are treated as a perturbation term.
This study suggests that applying magnetic fields induces excitonic and LS-HS-ordered phases from the LS phase.
Additionally, another excitonic phase associated with the LS-HS order is found between the LS-HS and HS phases in a lower-field region.
This phase is believed to correspond to the ESS shown in Fig.~\ref{J_h}(a).
Note that the strength of external magnetic fields inducing excitonic states in the strong correlation limit is significantly lower than that obtained by the Hartree-Fock approximation.
This discrepancy is attributed to the overestimation of on-site interactions within this approximation.
When electron correlations are more precisely considered beyond the Hartree-Fock theory, the critical field required to induce excitonic states is expected to be lower.

Recent experimental studies on cobalt oxides have suggested the presence of two types of excitonic states under ultra-high magnetic fields, reaching up to 600~T~\cite{Ikeda2023}.
Moreover, theoretical analysis of the localized model within the strong correlation limit, which is based on the five-orbital Hubbard model for the $3d$ orbitals, has discussed the possibility that these states are an excitonic condensation of the IS states and a supersolid. 
This is consistent with the findings of this study, which employs an approach based on weak correlations in the two-orbital Hubbard model.
Hence, it is inferred that, even beyond the strong correlation regime, the five-orbital Hubbard model could support not only excitonic states but also supersolid states.

Here, we compare the magnitude of the critical magnetic field observed experimentally with the predictions of our study.
The transfer integral $t_a$ represents the electron hopping between $e_g$ orbitals via $p$ orbitals of an oxygen ion, typically estimated to be $t_a \sim 1~\mathrm{eV}$~\cite{Sotnikov2016, Mizokawa1996, Saitoh1997}.
Based on this consideration, our results suggest that the critical magnetic field for the transition from the LS phase to the EC phase is $\sim 1000~\mathrm{T}$.
On the other hand, field-induced phase transitions in LaCoO$_3$ have been experimentally observed at approximately $70$~T at 5~K and $170$~T at 78~K, significantly lower than the predictions of the current calculations. 
We attribute this discrepancy to the limitations of the Hartree-Fock approximation, which tends to overestimate the effects of the Coulomb interactions, as previously discussed.
Properly incorporating many-body effects is expected to reduce the required field intensity inducing the EC phase from the LS phase.

Finally, we address how the effects of spin-orbit coupling and orbital magnetic moments manifest in real materials.
Let us consider the case where the $d_{x^2-y^2}$ and $d_{xy}$ orbitals are hybridized to realize the EC phase. As indicated in Ref.~\cite{Nasu2020}, it is expected that lattice distortions accompany this EC phase because the $z$-axis becomes inequivalent to the other two axes.
Indeed, in $\ce{Pr_{0.5}Ca_{0.5}CoO3}$, lattice distortions accompanying a phase transition at $T_s \sim 90$~K have been observed~\cite{Tsubouchi2002, Fujita2004}.
On the other hand, no magnetic order below $T_s$ has been reported in this material.
Our calculation results with spin-orbit coupling [Fig.~\ref{l_J}(e)] indicate that the EC phase accompanies a magnetic ordering, which appears to be inconsistent with the experimental results.
Nevertheless, the experimental results have also reported the presence of a small spin splitting below the transition temperature~\cite{Hejtmanek2013, Hejtmanek2010}, which may not contradict the presence of magnetic order with very small ordered moments obtained in this study.
Furthermore, experimental studies have also been conducted on epitaxially grown $\ce{LaCoO3}$ thin films on substrates, which enables applying pseudo pressure in a specific direction to control the spin and orbital states~\cite{Fujioka2013, Fujioka2015, Yamasaki2016, Yokoyama2018}.
In this context, the realization of the EC phase through hybridization between $d_{x^2-y^2}$ and $d_{xy}$ is expected due to the two-dimensionality of the system.
The superposition of these two orbitals can create significant orbital angular momentum in the out-of-plane direction, which could strongly influence the magnetism due to the orbital magnetic moment in the presence of a magnetic field.
Consequently, reducing the dimensionality of cobalt oxides might make it feasible to experimentally observe the directional dependence of applied magnetic fields due to spin-orbit coupling and orbital magnetic moments, as demonstrated in the present calculations.

\section{Summary}
\label{sec:summary}

In summary, we have examined the two-orbital Hubbard model with a crystalline field splitting under magnetic fields, employing the Hartree-Fock approximation.
Our findings reveal that when spin-orbit coupling and orbital magnetic moments are omitted, applying magnetic fields to the low-spin state induces spin-triplet excitonic and excitonic supersolid states.
The latter is characterized by the coexistence of spin-triplet excitonic and a staggered-type spin-state order.
Introducing spin-orbit coupling leads to an antiferromagnetic order with notably small ordered moments in the excitonic phase. 
Moreover, we have revealed that the Zeeman effect originating from the orbital magnetic moments destabilizes the spin-triplet excitonic state and leads to the low-spin state, which suggests that the orbital Zeeman effect brings about a substantial change to the magnetic phase diagram.
This phenomenon arises due to a substantial orbital moment generated by the $d_{x^2-y^2}$ and $d_{xy}$ orbitals, which are considered active in excitonic condensation in perovskite cobalt oxides.

\begin{acknowledgments}
The authors thank M.~Naka, A.~Ono, and A.~Hariki for fruitful discussions.
Parts of the numerical calculations were performed in the supercomputing
systems in ISSP, the University of Tokyo.
This work was supported by Grant-in-Aid for Scientific Research from JSPS, KAKENHI Grant No.~JP23H04859, JP23H04865, and JP23H01129.
\end{acknowledgments}

\bibliography{./refs}

\begin{thebibliography}{53}%
\makeatletter
\providecommand \@ifxundefined [1]{%
 \@ifx{#1\undefined}
}%
\providecommand \@ifnum [1]{%
 \ifnum #1\expandafter \@firstoftwo
 \else \expandafter \@secondoftwo
 \fi
}%
\providecommand \@ifx [1]{%
 \ifx #1\expandafter \@firstoftwo
 \else \expandafter \@secondoftwo
 \fi
}%
\providecommand \natexlab [1]{#1}%
\providecommand \enquote  [1]{``#1''}%
\providecommand \bibnamefont  [1]{#1}%
\providecommand \bibfnamefont [1]{#1}%
\providecommand \citenamefont [1]{#1}%
\providecommand \href@noop [0]{\@secondoftwo}%
\providecommand \href [0]{\begingroup \@sanitize@url \@href}%
\providecommand \@href[1]{\@@startlink{#1}\@@href}%
\providecommand \@@href[1]{\endgroup#1\@@endlink}%
\providecommand \@sanitize@url [0]{\catcode `\\12\catcode `\$12\catcode
  `\&12\catcode `\#12\catcode `\^12\catcode `\_12\catcode `\%12\relax}%
\providecommand \@@startlink[1]{}%
\providecommand \@@endlink[0]{}%
\providecommand \url  [0]{\begingroup\@sanitize@url \@url }%
\providecommand \@url [1]{\endgroup\@href {#1}{\urlprefix }}%
\providecommand \urlprefix  [0]{URL }%
\providecommand \Eprint [0]{\href }%
\providecommand \doibase [0]{http://dx.doi.org/}%
\providecommand \selectlanguage [0]{\@gobble}%
\providecommand \bibinfo  [0]{\@secondoftwo}%
\providecommand \bibfield  [0]{\@secondoftwo}%
\providecommand \translation [1]{[#1]}%
\providecommand \BibitemOpen [0]{}%
\providecommand \bibitemStop [0]{}%
\providecommand \bibitemNoStop [0]{.\EOS\space}%
\providecommand \EOS [0]{\spacefactor3000\relax}%
\providecommand \BibitemShut  [1]{\csname bibitem#1\endcsname}%
\let\auto@bib@innerbib\@empty
%</preamble>
\bibitem [{\citenamefont {Mott}(1961)}]{Mott1961}%
  \BibitemOpen
  \bibfield  {author} {\bibinfo {author} {\bibfnamefont {N.~F.}\ \bibnamefont
  {Mott}},\ }\href {\doibase 10.1080/14786436108243318} {\bibfield  {journal}
  {\bibinfo  {journal} {Philos. Mag.}\ }\textbf {\bibinfo {volume} {6}},\
  \bibinfo {pages} {287} (\bibinfo {year} {1961})}\BibitemShut {NoStop}%
\bibitem [{\citenamefont {J\'erome}\ \emph {et~al.}(1967)\citenamefont
  {J\'erome}, \citenamefont {Rice},\ and\ \citenamefont {Kohn}}]{Jerome1967}%
  \BibitemOpen
  \bibfield  {author} {\bibinfo {author} {\bibfnamefont {D.}~\bibnamefont
  {J\'erome}}, \bibinfo {author} {\bibfnamefont {T.~M.}\ \bibnamefont {Rice}},
  \ and\ \bibinfo {author} {\bibfnamefont {W.}~\bibnamefont {Kohn}},\ }\href
  {\doibase 10.1103/PhysRev.158.462} {\bibfield  {journal} {\bibinfo  {journal}
  {Phys. Rev.}\ }\textbf {\bibinfo {volume} {158}},\ \bibinfo {pages} {462}
  (\bibinfo {year} {1967})}\BibitemShut {NoStop}%
\bibitem [{\citenamefont {Halperin}\ and\ \citenamefont
  {Rice}(1968)}]{Halperin1968}%
  \BibitemOpen
  \bibfield  {author} {\bibinfo {author} {\bibfnamefont {B.~I.}\ \bibnamefont
  {Halperin}}\ and\ \bibinfo {author} {\bibfnamefont {T.~M.}\ \bibnamefont
  {Rice}},\ }\href {\doibase 10.1103/RevModPhys.40.755} {\bibfield  {journal}
  {\bibinfo  {journal} {Rev. Mod. Phys.}\ }\textbf {\bibinfo {volume} {40}},\
  \bibinfo {pages} {755} (\bibinfo {year} {1968})}\BibitemShut {NoStop}%
\bibitem [{\citenamefont {Wakisaka}\ \emph {et~al.}(2009)\citenamefont
  {Wakisaka}, \citenamefont {Sudayama}, \citenamefont {Takubo}, \citenamefont
  {Mizokawa}, \citenamefont {Arita}, \citenamefont {Namatame}, \citenamefont
  {Taniguchi}, \citenamefont {Katayama}, \citenamefont {Nohara},\ and\
  \citenamefont {Takagi}}]{Wakisaka2009}%
  \BibitemOpen
  \bibfield  {author} {\bibinfo {author} {\bibfnamefont {Y.}~\bibnamefont
  {Wakisaka}}, \bibinfo {author} {\bibfnamefont {T.}~\bibnamefont {Sudayama}},
  \bibinfo {author} {\bibfnamefont {K.}~\bibnamefont {Takubo}}, \bibinfo
  {author} {\bibfnamefont {T.}~\bibnamefont {Mizokawa}}, \bibinfo {author}
  {\bibfnamefont {M.}~\bibnamefont {Arita}}, \bibinfo {author} {\bibfnamefont
  {H.}~\bibnamefont {Namatame}}, \bibinfo {author} {\bibfnamefont
  {M.}~\bibnamefont {Taniguchi}}, \bibinfo {author} {\bibfnamefont
  {N.}~\bibnamefont {Katayama}}, \bibinfo {author} {\bibfnamefont
  {M.}~\bibnamefont {Nohara}}, \ and\ \bibinfo {author} {\bibfnamefont
  {H.}~\bibnamefont {Takagi}},\ }\href {\doibase
  10.1103/PhysRevLett.103.026402} {\bibfield  {journal} {\bibinfo  {journal}
  {Phys. Rev. Lett.}\ }\textbf {\bibinfo {volume} {103}},\ \bibinfo {pages}
  {026402} (\bibinfo {year} {2009})}\BibitemShut {NoStop}%
\bibitem [{\citenamefont {Seki}\ \emph {et~al.}(2014)\citenamefont {Seki},
  \citenamefont {Wakisaka}, \citenamefont {Kaneko}, \citenamefont {Toriyama},
  \citenamefont {Konishi}, \citenamefont {Sudayama}, \citenamefont {Saini},
  \citenamefont {Arita}, \citenamefont {Namatame}, \citenamefont {Taniguchi},
  \citenamefont {Katayama}, \citenamefont {Nohara}, \citenamefont {Takagi},
  \citenamefont {Mizokawa},\ and\ \citenamefont {Ohta}}]{Seki2014}%
  \BibitemOpen
  \bibfield  {author} {\bibinfo {author} {\bibfnamefont {K.}~\bibnamefont
  {Seki}}, \bibinfo {author} {\bibfnamefont {Y.}~\bibnamefont {Wakisaka}},
  \bibinfo {author} {\bibfnamefont {T.}~\bibnamefont {Kaneko}}, \bibinfo
  {author} {\bibfnamefont {T.}~\bibnamefont {Toriyama}}, \bibinfo {author}
  {\bibfnamefont {T.}~\bibnamefont {Konishi}}, \bibinfo {author} {\bibfnamefont
  {T.}~\bibnamefont {Sudayama}}, \bibinfo {author} {\bibfnamefont {N.~L.}\
  \bibnamefont {Saini}}, \bibinfo {author} {\bibfnamefont {M.}~\bibnamefont
  {Arita}}, \bibinfo {author} {\bibfnamefont {H.}~\bibnamefont {Namatame}},
  \bibinfo {author} {\bibfnamefont {M.}~\bibnamefont {Taniguchi}}, \bibinfo
  {author} {\bibfnamefont {N.}~\bibnamefont {Katayama}}, \bibinfo {author}
  {\bibfnamefont {M.}~\bibnamefont {Nohara}}, \bibinfo {author} {\bibfnamefont
  {H.}~\bibnamefont {Takagi}}, \bibinfo {author} {\bibfnamefont
  {T.}~\bibnamefont {Mizokawa}}, \ and\ \bibinfo {author} {\bibfnamefont
  {Y.}~\bibnamefont {Ohta}},\ }\href {\doibase 10.1103/PhysRevB.90.155116}
  {\bibfield  {journal} {\bibinfo  {journal} {Phys. Rev. B}\ }\textbf {\bibinfo
  {volume} {90}},\ \bibinfo {pages} {155116} (\bibinfo {year}
  {2014})}\BibitemShut {NoStop}%
\bibitem [{\citenamefont {{Di Salvo}}\ \emph {et~al.}(1986)\citenamefont {{Di
  Salvo}}, \citenamefont {Chen}, \citenamefont {Fleming}, \citenamefont
  {Waszczak}, \citenamefont {Dunn}, \citenamefont {Sunshine},\ and\
  \citenamefont {Ibers}}]{Salvo1986}%
  \BibitemOpen
  \bibfield  {author} {\bibinfo {author} {\bibfnamefont {F.}~\bibnamefont {{Di
  Salvo}}}, \bibinfo {author} {\bibfnamefont {C.}~\bibnamefont {Chen}},
  \bibinfo {author} {\bibfnamefont {R.}~\bibnamefont {Fleming}}, \bibinfo
  {author} {\bibfnamefont {J.}~\bibnamefont {Waszczak}}, \bibinfo {author}
  {\bibfnamefont {R.}~\bibnamefont {Dunn}}, \bibinfo {author} {\bibfnamefont
  {S.}~\bibnamefont {Sunshine}}, \ and\ \bibinfo {author} {\bibfnamefont
  {J.~A.}\ \bibnamefont {Ibers}},\ }\href {\doibase
  https://doi.org/10.1016/0022-5088(86)90216-X} {\bibfield  {journal} {\bibinfo
   {journal} {J. Less-Common Met.}\ }\textbf {\bibinfo {volume} {116}},\
  \bibinfo {pages} {51} (\bibinfo {year} {1986})}\BibitemShut {NoStop}%
\bibitem [{\citenamefont {Kaneko}\ \emph {et~al.}(2013)\citenamefont {Kaneko},
  \citenamefont {Toriyama}, \citenamefont {Konishi},\ and\ \citenamefont
  {Ohta}}]{Kaneko2013}%
  \BibitemOpen
  \bibfield  {author} {\bibinfo {author} {\bibfnamefont {T.}~\bibnamefont
  {Kaneko}}, \bibinfo {author} {\bibfnamefont {T.}~\bibnamefont {Toriyama}},
  \bibinfo {author} {\bibfnamefont {T.}~\bibnamefont {Konishi}}, \ and\
  \bibinfo {author} {\bibfnamefont {Y.}~\bibnamefont {Ohta}},\ }\href {\doibase
  10.1103/PhysRevB.87.035121} {\bibfield  {journal} {\bibinfo  {journal} {Phys.
  Rev. B}\ }\textbf {\bibinfo {volume} {87}},\ \bibinfo {pages} {035121}
  (\bibinfo {year} {2013})}\BibitemShut {NoStop}%
\bibitem [{\citenamefont {Kaneko}\ \emph {et~al.}(2015)\citenamefont {Kaneko},
  \citenamefont {Zenker}, \citenamefont {Fehske},\ and\ \citenamefont
  {Ohta}}]{Kaneko2015}%
  \BibitemOpen
  \bibfield  {author} {\bibinfo {author} {\bibfnamefont {T.}~\bibnamefont
  {Kaneko}}, \bibinfo {author} {\bibfnamefont {B.}~\bibnamefont {Zenker}},
  \bibinfo {author} {\bibfnamefont {H.}~\bibnamefont {Fehske}}, \ and\ \bibinfo
  {author} {\bibfnamefont {Y.}~\bibnamefont {Ohta}},\ }\href {\doibase
  10.1103/PhysRevB.92.115106} {\bibfield  {journal} {\bibinfo  {journal} {Phys.
  Rev. B}\ }\textbf {\bibinfo {volume} {92}},\ \bibinfo {pages} {115106}
  (\bibinfo {year} {2015})}\BibitemShut {NoStop}%
\bibitem [{\citenamefont {Kune\ifmmode~\check{s}\else \v{s}\fi{}}\ and\
  \citenamefont {Augustinsk\'y}(2014{\natexlab{a}})}]{Kunes2014condensation}%
  \BibitemOpen
  \bibfield  {author} {\bibinfo {author} {\bibfnamefont {J.}~\bibnamefont
  {Kune\ifmmode~\check{s}\else \v{s}\fi{}}}\ and\ \bibinfo {author}
  {\bibfnamefont {P.}~\bibnamefont {Augustinsk\'y}},\ }\href {\doibase
  10.1103/PhysRevB.90.235112} {\bibfield  {journal} {\bibinfo  {journal} {Phys.
  Rev. B}\ }\textbf {\bibinfo {volume} {90}},\ \bibinfo {pages} {235112}
  (\bibinfo {year} {2014}{\natexlab{a}})}\BibitemShut {NoStop}%
\bibitem [{\citenamefont {Kune\ifmmode~\check{s}\else \v{s}\fi{}}\ and\
  \citenamefont {Augustinsk\'y}(2014{\natexlab{b}})}]{Kunes2014instability}%
  \BibitemOpen
  \bibfield  {author} {\bibinfo {author} {\bibfnamefont {J.}~\bibnamefont
  {Kune\ifmmode~\check{s}\else \v{s}\fi{}}}\ and\ \bibinfo {author}
  {\bibfnamefont {P.}~\bibnamefont {Augustinsk\'y}},\ }\href {\doibase
  10.1103/PhysRevB.89.115134} {\bibfield  {journal} {\bibinfo  {journal} {Phys.
  Rev. B}\ }\textbf {\bibinfo {volume} {89}},\ \bibinfo {pages} {115134}
  (\bibinfo {year} {2014}{\natexlab{b}})}\BibitemShut {NoStop}%
\bibitem [{\citenamefont {Tokura}\ \emph {et~al.}(1998)\citenamefont {Tokura},
  \citenamefont {Okimoto}, \citenamefont {Yamaguchi}, \citenamefont
  {Taniguchi}, \citenamefont {Kimura},\ and\ \citenamefont
  {Takagi}}]{Tokura1998}%
  \BibitemOpen
  \bibfield  {author} {\bibinfo {author} {\bibfnamefont {Y.}~\bibnamefont
  {Tokura}}, \bibinfo {author} {\bibfnamefont {Y.}~\bibnamefont {Okimoto}},
  \bibinfo {author} {\bibfnamefont {S.}~\bibnamefont {Yamaguchi}}, \bibinfo
  {author} {\bibfnamefont {H.}~\bibnamefont {Taniguchi}}, \bibinfo {author}
  {\bibfnamefont {T.}~\bibnamefont {Kimura}}, \ and\ \bibinfo {author}
  {\bibfnamefont {H.}~\bibnamefont {Takagi}},\ }\href {\doibase
  10.1103/PhysRevB.58.R1699} {\bibfield  {journal} {\bibinfo  {journal} {Phys.
  Rev. B}\ }\textbf {\bibinfo {volume} {58}},\ \bibinfo {pages} {R1699}
  (\bibinfo {year} {1998})}\BibitemShut {NoStop}%
\bibitem [{\citenamefont {Asai}\ \emph {et~al.}(1998)\citenamefont {Asai},
  \citenamefont {Yoneda}, \citenamefont {Yokokura}, \citenamefont {Tranquada},
  \citenamefont {Shirane},\ and\ \citenamefont {Kohn}}]{Asai1998}%
  \BibitemOpen
  \bibfield  {author} {\bibinfo {author} {\bibfnamefont {K.}~\bibnamefont
  {Asai}}, \bibinfo {author} {\bibfnamefont {A.}~\bibnamefont {Yoneda}},
  \bibinfo {author} {\bibfnamefont {O.}~\bibnamefont {Yokokura}}, \bibinfo
  {author} {\bibfnamefont {J.}~\bibnamefont {Tranquada}}, \bibinfo {author}
  {\bibfnamefont {G.}~\bibnamefont {Shirane}}, \ and\ \bibinfo {author}
  {\bibfnamefont {K.}~\bibnamefont {Kohn}},\ }\href {\doibase
  10.1143/JPSJ.67.290} {\bibfield  {journal} {\bibinfo  {journal} {J. Phys.
  Soc. Jpn.}\ }\textbf {\bibinfo {volume} {67}},\ \bibinfo {pages} {290}
  (\bibinfo {year} {1998})}\BibitemShut {NoStop}%
\bibitem [{\citenamefont {Ikeda}\ \emph {et~al.}()\citenamefont {Ikeda},
  \citenamefont {Matsuda}, \citenamefont {Sato}, \citenamefont {Ishii},
  \citenamefont {Sawabe}, \citenamefont {Nakamura}, \citenamefont {Takeyama},\
  and\ \citenamefont {Nasu}}]{Ikeda2023}%
  \BibitemOpen
  \bibfield  {author} {\bibinfo {author} {\bibfnamefont {A.}~\bibnamefont
  {Ikeda}}, \bibinfo {author} {\bibfnamefont {Y.~H.}\ \bibnamefont {Matsuda}},
  \bibinfo {author} {\bibfnamefont {K.}~\bibnamefont {Sato}}, \bibinfo {author}
  {\bibfnamefont {Y.}~\bibnamefont {Ishii}}, \bibinfo {author} {\bibfnamefont
  {H.}~\bibnamefont {Sawabe}}, \bibinfo {author} {\bibfnamefont
  {D.}~\bibnamefont {Nakamura}}, \bibinfo {author} {\bibfnamefont
  {S.}~\bibnamefont {Takeyama}}, \ and\ \bibinfo {author} {\bibfnamefont
  {J.}~\bibnamefont {Nasu}},\ }\href {\doibase 10.1038/s41467-023-37125-4}
  {\bibfield  {journal} {\bibinfo  {journal} {Nat. Commun.}\ }\textbf {\bibinfo
  {volume} {14}},\ \bibinfo {pages} {1744}}\BibitemShut {NoStop}%
\bibitem [{\citenamefont {Nasu}\ \emph {et~al.}(2016)\citenamefont {Nasu},
  \citenamefont {Watanabe}, \citenamefont {Naka},\ and\ \citenamefont
  {Ishihara}}]{Nasu2016}%
  \BibitemOpen
  \bibfield  {author} {\bibinfo {author} {\bibfnamefont {J.}~\bibnamefont
  {Nasu}}, \bibinfo {author} {\bibfnamefont {T.}~\bibnamefont {Watanabe}},
  \bibinfo {author} {\bibfnamefont {M.}~\bibnamefont {Naka}}, \ and\ \bibinfo
  {author} {\bibfnamefont {S.}~\bibnamefont {Ishihara}},\ }\href {\doibase
  10.1103/PhysRevB.93.205136} {\bibfield  {journal} {\bibinfo  {journal} {Phys.
  Rev. B}\ }\textbf {\bibinfo {volume} {93}},\ \bibinfo {pages} {205136}
  (\bibinfo {year} {2016})}\BibitemShut {NoStop}%
\bibitem [{\citenamefont {Kaneko}\ and\ \citenamefont
  {Ohta}(2014)}]{Kaneko2014}%
  \BibitemOpen
  \bibfield  {author} {\bibinfo {author} {\bibfnamefont {T.}~\bibnamefont
  {Kaneko}}\ and\ \bibinfo {author} {\bibfnamefont {Y.}~\bibnamefont {Ohta}},\
  }\href {\doibase 10.1103/PhysRevB.90.245144} {\bibfield  {journal} {\bibinfo
  {journal} {Phys. Rev. B}\ }\textbf {\bibinfo {volume} {90}},\ \bibinfo
  {pages} {245144} (\bibinfo {year} {2014})}\BibitemShut {NoStop}%
\bibitem [{\citenamefont {Tsubouchi}\ \emph {et~al.}(2002)\citenamefont
  {Tsubouchi}, \citenamefont {Ky\^omen}, \citenamefont {Itoh}, \citenamefont
  {Ganguly}, \citenamefont {Oguni}, \citenamefont {Shimojo}, \citenamefont
  {Morii},\ and\ \citenamefont {Ishii}}]{Tsubouchi2002}%
  \BibitemOpen
  \bibfield  {author} {\bibinfo {author} {\bibfnamefont {S.}~\bibnamefont
  {Tsubouchi}}, \bibinfo {author} {\bibfnamefont {T.}~\bibnamefont {Ky\^omen}},
  \bibinfo {author} {\bibfnamefont {M.}~\bibnamefont {Itoh}}, \bibinfo {author}
  {\bibfnamefont {P.}~\bibnamefont {Ganguly}}, \bibinfo {author} {\bibfnamefont
  {M.}~\bibnamefont {Oguni}}, \bibinfo {author} {\bibfnamefont
  {Y.}~\bibnamefont {Shimojo}}, \bibinfo {author} {\bibfnamefont
  {Y.}~\bibnamefont {Morii}}, \ and\ \bibinfo {author} {\bibfnamefont
  {Y.}~\bibnamefont {Ishii}},\ }\href {\doibase 10.1103/PhysRevB.66.052418}
  {\bibfield  {journal} {\bibinfo  {journal} {Phys. Rev. B}\ }\textbf {\bibinfo
  {volume} {66}},\ \bibinfo {pages} {052418} (\bibinfo {year}
  {2002})}\BibitemShut {NoStop}%
\bibitem [{\citenamefont {Fujita}\ \emph {et~al.}(2004)\citenamefont {Fujita},
  \citenamefont {Miyashita}, \citenamefont {Yasui}, \citenamefont {Kobayashi},
  \citenamefont {Sato}, \citenamefont {Nishibori}, \citenamefont {Sakata},
  \citenamefont {Shimojo}, \citenamefont {Igawa}, \citenamefont {Ishii},
  \citenamefont {Kakurai}, \citenamefont {Adachi}, \citenamefont {Ohishi},\
  and\ \citenamefont {Takata}}]{Fujita2004}%
  \BibitemOpen
  \bibfield  {author} {\bibinfo {author} {\bibfnamefont {T.}~\bibnamefont
  {Fujita}}, \bibinfo {author} {\bibfnamefont {T.}~\bibnamefont {Miyashita}},
  \bibinfo {author} {\bibfnamefont {Y.}~\bibnamefont {Yasui}}, \bibinfo
  {author} {\bibfnamefont {Y.}~\bibnamefont {Kobayashi}}, \bibinfo {author}
  {\bibfnamefont {M.}~\bibnamefont {Sato}}, \bibinfo {author} {\bibfnamefont
  {E.}~\bibnamefont {Nishibori}}, \bibinfo {author} {\bibfnamefont
  {M.}~\bibnamefont {Sakata}}, \bibinfo {author} {\bibfnamefont
  {Y.}~\bibnamefont {Shimojo}}, \bibinfo {author} {\bibfnamefont
  {N.}~\bibnamefont {Igawa}}, \bibinfo {author} {\bibfnamefont
  {Y.}~\bibnamefont {Ishii}}, \bibinfo {author} {\bibfnamefont
  {K.}~\bibnamefont {Kakurai}}, \bibinfo {author} {\bibfnamefont
  {T.}~\bibnamefont {Adachi}}, \bibinfo {author} {\bibfnamefont
  {Y.}~\bibnamefont {Ohishi}}, \ and\ \bibinfo {author} {\bibfnamefont
  {M.}~\bibnamefont {Takata}},\ }\href {\doibase 10.1143/JPSJ.73.1987}
  {\bibfield  {journal} {\bibinfo  {journal} {J. Phys. Soc. Jpn.}\ }\textbf
  {\bibinfo {volume} {73}},\ \bibinfo {pages} {1987} (\bibinfo {year}
  {2004})}\BibitemShut {NoStop}%
\bibitem [{\citenamefont {Hejtm^^c3^^a1nek}\ \emph {et~al.}(2013)\citenamefont
  {Hejtm^^c3^^a1nek}, \citenamefont {Jir^^c3^^a1k}, \citenamefont {Kaman},
  \citenamefont {Kn^^c3^^ad^^c5^^beek}, \citenamefont {^^c5^^a0antav^^c3^^a1},
  \citenamefont {Nitta}, \citenamefont {Naito},\ and\ \citenamefont
  {Fujishiro}}]{Hejtmanek2013}%
  \BibitemOpen
  \bibfield  {author} {\bibinfo {author} {\bibfnamefont {J.}~\bibnamefont
  {Hejtm^^c3^^a1nek}}, \bibinfo {author} {\bibfnamefont {Z.}~\bibnamefont
  {Jir^^c3^^a1k}}, \bibinfo {author} {\bibfnamefont {O.}~\bibnamefont {Kaman}},
  \bibinfo {author} {\bibfnamefont {K.}~\bibnamefont {Kn^^c3^^ad^^c5^^beek}},
  \bibinfo {author} {\bibfnamefont {E.}~\bibnamefont {^^c5^^a0antav^^c3^^a1}},
  \bibinfo {author} {\bibfnamefont {K.}~\bibnamefont {Nitta}}, \bibinfo
  {author} {\bibfnamefont {T.}~\bibnamefont {Naito}}, \ and\ \bibinfo {author}
  {\bibfnamefont {H.}~\bibnamefont {Fujishiro}},\ }\href {\doibase
  10.1140/epjb/e2013-30653-y} {\bibfield  {journal} {\bibinfo  {journal} {Eur.
  Phys. J. B}\ }\textbf {\bibinfo {volume} {86}},\ \bibinfo {pages} {305}
  (\bibinfo {year} {2013})}\BibitemShut {NoStop}%
\bibitem [{\citenamefont {Kaneko}\ and\ \citenamefont
  {Ohta}(2016)}]{Kanekomultiploe2016}%
  \BibitemOpen
  \bibfield  {author} {\bibinfo {author} {\bibfnamefont {T.}~\bibnamefont
  {Kaneko}}\ and\ \bibinfo {author} {\bibfnamefont {Y.}~\bibnamefont {Ohta}},\
  }\href {\doibase 10.1103/PhysRevB.94.125127} {\bibfield  {journal} {\bibinfo
  {journal} {Phys. Rev. B}\ }\textbf {\bibinfo {volume} {94}},\ \bibinfo
  {pages} {125127} (\bibinfo {year} {2016})}\BibitemShut {NoStop}%
\bibitem [{\citenamefont {Yamaguchi}\ \emph {et~al.}(2017)\citenamefont
  {Yamaguchi}, \citenamefont {Sugimoto},\ and\ \citenamefont
  {Ohta}}]{Yamaguchi2017}%
  \BibitemOpen
  \bibfield  {author} {\bibinfo {author} {\bibfnamefont {T.}~\bibnamefont
  {Yamaguchi}}, \bibinfo {author} {\bibfnamefont {K.}~\bibnamefont {Sugimoto}},
  \ and\ \bibinfo {author} {\bibfnamefont {Y.}~\bibnamefont {Ohta}},\ }\href
  {\doibase 10.7566/JPSJ.86.043701} {\bibfield  {journal} {\bibinfo  {journal}
  {J. Phys. Soc. Jpn.}\ }\textbf {\bibinfo {volume} {86}},\ \bibinfo {pages}
  {043701} (\bibinfo {year} {2017})}\BibitemShut {NoStop}%
\bibitem [{\citenamefont {Altarawneh}\ \emph {et~al.}(2012)\citenamefont
  {Altarawneh}, \citenamefont {Chern}, \citenamefont {Harrison}, \citenamefont
  {Batista}, \citenamefont {Uchida}, \citenamefont {Jaime}, \citenamefont
  {Rickel}, \citenamefont {Crooker}, \citenamefont {Mielke}, \citenamefont
  {Betts}, \citenamefont {Mitchell},\ and\ \citenamefont
  {Hoch}}]{Altarawneh2012}%
  \BibitemOpen
  \bibfield  {author} {\bibinfo {author} {\bibfnamefont {M.~M.}\ \bibnamefont
  {Altarawneh}}, \bibinfo {author} {\bibfnamefont {G.-W.}\ \bibnamefont
  {Chern}}, \bibinfo {author} {\bibfnamefont {N.}~\bibnamefont {Harrison}},
  \bibinfo {author} {\bibfnamefont {C.~D.}\ \bibnamefont {Batista}}, \bibinfo
  {author} {\bibfnamefont {A.}~\bibnamefont {Uchida}}, \bibinfo {author}
  {\bibfnamefont {M.}~\bibnamefont {Jaime}}, \bibinfo {author} {\bibfnamefont
  {D.~G.}\ \bibnamefont {Rickel}}, \bibinfo {author} {\bibfnamefont {S.~A.}\
  \bibnamefont {Crooker}}, \bibinfo {author} {\bibfnamefont {C.~H.}\
  \bibnamefont {Mielke}}, \bibinfo {author} {\bibfnamefont {J.~B.}\
  \bibnamefont {Betts}}, \bibinfo {author} {\bibfnamefont {J.~F.}\ \bibnamefont
  {Mitchell}}, \ and\ \bibinfo {author} {\bibfnamefont {M.~J.~R.}\ \bibnamefont
  {Hoch}},\ }\href {\doibase 10.1103/PhysRevLett.109.037201} {\bibfield
  {journal} {\bibinfo  {journal} {Phys. Rev. Lett.}\ }\textbf {\bibinfo
  {volume} {109}},\ \bibinfo {pages} {037201} (\bibinfo {year}
  {2012})}\BibitemShut {NoStop}%
\bibitem [{\citenamefont {Rotter}\ \emph {et~al.}(2014)\citenamefont {Rotter},
  \citenamefont {Wang}, \citenamefont {Boothroyd}, \citenamefont {Prabhakaran},
  \citenamefont {Tanaka},\ and\ \citenamefont {Doerr}}]{Rotter2014}%
  \BibitemOpen
  \bibfield  {author} {\bibinfo {author} {\bibfnamefont {M.}~\bibnamefont
  {Rotter}}, \bibinfo {author} {\bibfnamefont {Z.-S.}\ \bibnamefont {Wang}},
  \bibinfo {author} {\bibfnamefont {A.~T.}\ \bibnamefont {Boothroyd}}, \bibinfo
  {author} {\bibfnamefont {D.}~\bibnamefont {Prabhakaran}}, \bibinfo {author}
  {\bibfnamefont {A.}~\bibnamefont {Tanaka}}, \ and\ \bibinfo {author}
  {\bibfnamefont {M.}~\bibnamefont {Doerr}},\ }\href {\doibase
  10.1038/srep07003} {\bibfield  {journal} {\bibinfo  {journal} {Sci. Rep.}\
  }\textbf {\bibinfo {volume} {4}},\ \bibinfo {pages} {7003} (\bibinfo {year}
  {2014})}\BibitemShut {NoStop}%
\bibitem [{\citenamefont {Ikeda}\ \emph {et~al.}(2016)\citenamefont {Ikeda},
  \citenamefont {Nomura}, \citenamefont {Matsuda}, \citenamefont {Matsuo},
  \citenamefont {Kindo},\ and\ \citenamefont {Sato}}]{Ikeda2016}%
  \BibitemOpen
  \bibfield  {author} {\bibinfo {author} {\bibfnamefont {A.}~\bibnamefont
  {Ikeda}}, \bibinfo {author} {\bibfnamefont {T.}~\bibnamefont {Nomura}},
  \bibinfo {author} {\bibfnamefont {Y.~H.}\ \bibnamefont {Matsuda}}, \bibinfo
  {author} {\bibfnamefont {A.}~\bibnamefont {Matsuo}}, \bibinfo {author}
  {\bibfnamefont {K.}~\bibnamefont {Kindo}}, \ and\ \bibinfo {author}
  {\bibfnamefont {K.}~\bibnamefont {Sato}},\ }\href {\doibase
  10.1103/PhysRevB.93.220401} {\bibfield  {journal} {\bibinfo  {journal} {Phys.
  Rev. B}\ }\textbf {\bibinfo {volume} {93}},\ \bibinfo {pages} {220401}
  (\bibinfo {year} {2016})}\BibitemShut {NoStop}%
\bibitem [{\citenamefont {Ikeda}\ \emph {et~al.}(2020)\citenamefont {Ikeda},
  \citenamefont {Matsuda},\ and\ \citenamefont {Sato}}]{Ikeda2020}%
  \BibitemOpen
  \bibfield  {author} {\bibinfo {author} {\bibfnamefont {A.}~\bibnamefont
  {Ikeda}}, \bibinfo {author} {\bibfnamefont {Y.~H.}\ \bibnamefont {Matsuda}},
  \ and\ \bibinfo {author} {\bibfnamefont {K.}~\bibnamefont {Sato}},\ }\href
  {\doibase 10.1103/PhysRevLett.125.177202} {\bibfield  {journal} {\bibinfo
  {journal} {Phys. Rev. Lett.}\ }\textbf {\bibinfo {volume} {125}},\ \bibinfo
  {pages} {177202} (\bibinfo {year} {2020})}\BibitemShut {NoStop}%
\bibitem [{\citenamefont {Tatsuno}\ \emph {et~al.}(2016)\citenamefont
  {Tatsuno}, \citenamefont {Mizoguchi}, \citenamefont {Nasu}, \citenamefont
  {Naka},\ and\ \citenamefont {Ishihara}}]{Tatsuno2016}%
  \BibitemOpen
  \bibfield  {author} {\bibinfo {author} {\bibfnamefont {T.}~\bibnamefont
  {Tatsuno}}, \bibinfo {author} {\bibfnamefont {E.}~\bibnamefont {Mizoguchi}},
  \bibinfo {author} {\bibfnamefont {J.}~\bibnamefont {Nasu}}, \bibinfo {author}
  {\bibfnamefont {M.}~\bibnamefont {Naka}}, \ and\ \bibinfo {author}
  {\bibfnamefont {S.}~\bibnamefont {Ishihara}},\ }\href {\doibase
  10.7566/JPSJ.85.083706} {\bibfield  {journal} {\bibinfo  {journal} {J. Phys.
  Soc. Jpn.}\ }\textbf {\bibinfo {volume} {85}},\ \bibinfo {pages} {083706}
  (\bibinfo {year} {2016})}\BibitemShut {NoStop}%
\bibitem [{\citenamefont {Sotnikov}\ and\ \citenamefont
  {Kune^^c5^^a1}(2016)}]{Sotnikov2016}%
  \BibitemOpen
  \bibfield  {author} {\bibinfo {author} {\bibfnamefont {A.}~\bibnamefont
  {Sotnikov}}\ and\ \bibinfo {author} {\bibfnamefont {J.}~\bibnamefont
  {Kune^^c5^^a1}},\ }\href {\doibase 10.1038/srep30510} {\bibfield  {journal}
  {\bibinfo  {journal} {Sci. Rep.}\ }\textbf {\bibinfo {volume} {6}},\ \bibinfo
  {pages} {30510} (\bibinfo {year} {2016})}\BibitemShut {NoStop}%
\bibitem [{\citenamefont {Kitagawa}\ and\ \citenamefont
  {Matsueda}(2022)}]{Kitagawa2022}%
  \BibitemOpen
  \bibfield  {author} {\bibinfo {author} {\bibfnamefont {K.}~\bibnamefont
  {Kitagawa}}\ and\ \bibinfo {author} {\bibfnamefont {H.}~\bibnamefont
  {Matsueda}},\ }\href {\doibase 10.7566/JPSJ.91.104705} {\bibfield  {journal}
  {\bibinfo  {journal} {J. Phys. Soc. Jpn.}\ }\textbf {\bibinfo {volume}
  {91}},\ \bibinfo {pages} {104705} (\bibinfo {year} {2022})}\BibitemShut
  {NoStop}%
\bibitem [{\citenamefont {Kanamori}(1957{\natexlab{a}})}]{Kanamori1957_1}%
  \BibitemOpen
  \bibfield  {author} {\bibinfo {author} {\bibfnamefont {J.}~\bibnamefont
  {Kanamori}},\ }\href {\doibase 10.1143/PTP.17.177} {\bibfield  {journal}
  {\bibinfo  {journal} {Prog. Theor. Phys.}\ }\textbf {\bibinfo {volume}
  {17}},\ \bibinfo {pages} {177} (\bibinfo {year}
  {1957}{\natexlab{a}})}\BibitemShut {NoStop}%
\bibitem [{\citenamefont {Kanamori}(1957{\natexlab{b}})}]{Kanamori1957_2}%
  \BibitemOpen
  \bibfield  {author} {\bibinfo {author} {\bibfnamefont {J.}~\bibnamefont
  {Kanamori}},\ }\href {\doibase 10.1143/PTP.17.197} {\bibfield  {journal}
  {\bibinfo  {journal} {Prog. Theor. Phys.}\ }\textbf {\bibinfo {volume}
  {17}},\ \bibinfo {pages} {197} (\bibinfo {year}
  {1957}{\natexlab{b}})}\BibitemShut {NoStop}%
\bibitem [{\citenamefont {Tomiyasu}\ \emph {et~al.}(2011)\citenamefont
  {Tomiyasu}, \citenamefont {Crawford}, \citenamefont {Adroja}, \citenamefont
  {Manuel}, \citenamefont {Tominaga}, \citenamefont {Hara}, \citenamefont
  {Sato}, \citenamefont {Watanabe}, \citenamefont {Ikeda}, \citenamefont
  {Lynn}, \citenamefont {Iwasa},\ and\ \citenamefont {Yamada}}]{Tomiyasu2011}%
  \BibitemOpen
  \bibfield  {author} {\bibinfo {author} {\bibfnamefont {K.}~\bibnamefont
  {Tomiyasu}}, \bibinfo {author} {\bibfnamefont {M.~K.}\ \bibnamefont
  {Crawford}}, \bibinfo {author} {\bibfnamefont {D.~T.}\ \bibnamefont
  {Adroja}}, \bibinfo {author} {\bibfnamefont {P.}~\bibnamefont {Manuel}},
  \bibinfo {author} {\bibfnamefont {A.}~\bibnamefont {Tominaga}}, \bibinfo
  {author} {\bibfnamefont {S.}~\bibnamefont {Hara}}, \bibinfo {author}
  {\bibfnamefont {H.}~\bibnamefont {Sato}}, \bibinfo {author} {\bibfnamefont
  {T.}~\bibnamefont {Watanabe}}, \bibinfo {author} {\bibfnamefont {S.~I.}\
  \bibnamefont {Ikeda}}, \bibinfo {author} {\bibfnamefont {J.~W.}\ \bibnamefont
  {Lynn}}, \bibinfo {author} {\bibfnamefont {K.}~\bibnamefont {Iwasa}}, \ and\
  \bibinfo {author} {\bibfnamefont {K.}~\bibnamefont {Yamada}},\ }\href
  {\doibase 10.1103/PhysRevB.84.054405} {\bibfield  {journal} {\bibinfo
  {journal} {Phys. Rev. B}\ }\textbf {\bibinfo {volume} {84}},\ \bibinfo
  {pages} {054405} (\bibinfo {year} {2011})}\BibitemShut {NoStop}%
\bibitem [{\citenamefont {Haverkort}\ \emph {et~al.}(2006)\citenamefont
  {Haverkort}, \citenamefont {Hu}, \citenamefont {Cezar}, \citenamefont
  {Burnus}, \citenamefont {Hartmann}, \citenamefont {Reuther}, \citenamefont
  {Zobel}, \citenamefont {Lorenz}, \citenamefont {Tanaka}, \citenamefont
  {Brookes}, \citenamefont {Hsieh}, \citenamefont {Lin}, \citenamefont {Chen},\
  and\ \citenamefont {Tjeng}}]{Haverkort2006}%
  \BibitemOpen
  \bibfield  {author} {\bibinfo {author} {\bibfnamefont {M.~W.}\ \bibnamefont
  {Haverkort}}, \bibinfo {author} {\bibfnamefont {Z.}~\bibnamefont {Hu}},
  \bibinfo {author} {\bibfnamefont {J.~C.}\ \bibnamefont {Cezar}}, \bibinfo
  {author} {\bibfnamefont {T.}~\bibnamefont {Burnus}}, \bibinfo {author}
  {\bibfnamefont {H.}~\bibnamefont {Hartmann}}, \bibinfo {author}
  {\bibfnamefont {M.}~\bibnamefont {Reuther}}, \bibinfo {author} {\bibfnamefont
  {C.}~\bibnamefont {Zobel}}, \bibinfo {author} {\bibfnamefont
  {T.}~\bibnamefont {Lorenz}}, \bibinfo {author} {\bibfnamefont
  {A.}~\bibnamefont {Tanaka}}, \bibinfo {author} {\bibfnamefont {N.~B.}\
  \bibnamefont {Brookes}}, \bibinfo {author} {\bibfnamefont {H.~H.}\
  \bibnamefont {Hsieh}}, \bibinfo {author} {\bibfnamefont {H.-J.}\ \bibnamefont
  {Lin}}, \bibinfo {author} {\bibfnamefont {C.~T.}\ \bibnamefont {Chen}}, \
  and\ \bibinfo {author} {\bibfnamefont {L.~H.}\ \bibnamefont {Tjeng}},\ }\href
  {\doibase 10.1103/PhysRevLett.97.176405} {\bibfield  {journal} {\bibinfo
  {journal} {Phys. Rev. Lett.}\ }\textbf {\bibinfo {volume} {97}},\ \bibinfo
  {pages} {176405} (\bibinfo {year} {2006})}\BibitemShut {NoStop}%
\bibitem [{\citenamefont {Tomiyasu}\ \emph {et~al.}(2017)\citenamefont
  {Tomiyasu}, \citenamefont {Okamoto}, \citenamefont {Huang}, \citenamefont
  {Chen}, \citenamefont {Sinaga}, \citenamefont {Wu}, \citenamefont {Chu},
  \citenamefont {Singh}, \citenamefont {Wang}, \citenamefont {de~Groot},
  \citenamefont {Chainani}, \citenamefont {Ishihara}, \citenamefont {Chen},\
  and\ \citenamefont {Huang}}]{Tomiyasu2017}%
  \BibitemOpen
  \bibfield  {author} {\bibinfo {author} {\bibfnamefont {K.}~\bibnamefont
  {Tomiyasu}}, \bibinfo {author} {\bibfnamefont {J.}~\bibnamefont {Okamoto}},
  \bibinfo {author} {\bibfnamefont {H.~Y.}\ \bibnamefont {Huang}}, \bibinfo
  {author} {\bibfnamefont {Z.~Y.}\ \bibnamefont {Chen}}, \bibinfo {author}
  {\bibfnamefont {E.~P.}\ \bibnamefont {Sinaga}}, \bibinfo {author}
  {\bibfnamefont {W.~B.}\ \bibnamefont {Wu}}, \bibinfo {author} {\bibfnamefont
  {Y.~Y.}\ \bibnamefont {Chu}}, \bibinfo {author} {\bibfnamefont
  {A.}~\bibnamefont {Singh}}, \bibinfo {author} {\bibfnamefont {R.-P.}\
  \bibnamefont {Wang}}, \bibinfo {author} {\bibfnamefont {F.~M.~F.}\
  \bibnamefont {de~Groot}}, \bibinfo {author} {\bibfnamefont {A.}~\bibnamefont
  {Chainani}}, \bibinfo {author} {\bibfnamefont {S.}~\bibnamefont {Ishihara}},
  \bibinfo {author} {\bibfnamefont {C.~T.}\ \bibnamefont {Chen}}, \ and\
  \bibinfo {author} {\bibfnamefont {D.~J.}\ \bibnamefont {Huang}},\ }\href
  {\doibase 10.1103/PhysRevLett.119.196402} {\bibfield  {journal} {\bibinfo
  {journal} {Phys. Rev. Lett.}\ }\textbf {\bibinfo {volume} {119}},\ \bibinfo
  {pages} {196402} (\bibinfo {year} {2017})}\BibitemShut {NoStop}%
\bibitem [{\citenamefont {Nasu}\ \emph {et~al.}(2020)\citenamefont {Nasu},
  \citenamefont {Naka},\ and\ \citenamefont {Ishihara}}]{Nasu2020}%
  \BibitemOpen
  \bibfield  {author} {\bibinfo {author} {\bibfnamefont {J.}~\bibnamefont
  {Nasu}}, \bibinfo {author} {\bibfnamefont {M.}~\bibnamefont {Naka}}, \ and\
  \bibinfo {author} {\bibfnamefont {S.}~\bibnamefont {Ishihara}},\ }\href
  {\doibase 10.1103/PhysRevB.102.045143} {\bibfield  {journal} {\bibinfo
  {journal} {Phys. Rev. B}\ }\textbf {\bibinfo {volume} {102}},\ \bibinfo
  {pages} {045143} (\bibinfo {year} {2020})}\BibitemShut {NoStop}%
\bibitem [{\citenamefont {Werner}\ and\ \citenamefont
  {Millis}(2007)}]{Werner2007}%
  \BibitemOpen
  \bibfield  {author} {\bibinfo {author} {\bibfnamefont {P.}~\bibnamefont
  {Werner}}\ and\ \bibinfo {author} {\bibfnamefont {A.~J.}\ \bibnamefont
  {Millis}},\ }\href {\doibase 10.1103/PhysRevLett.99.126405} {\bibfield
  {journal} {\bibinfo  {journal} {Phys. Rev. Lett.}\ }\textbf {\bibinfo
  {volume} {99}},\ \bibinfo {pages} {126405} (\bibinfo {year}
  {2007})}\BibitemShut {NoStop}%
\bibitem [{\citenamefont {Suzuki}\ \emph {et~al.}(2009)\citenamefont {Suzuki},
  \citenamefont {Watanabe},\ and\ \citenamefont {Ishihara}}]{Suzuki2009}%
  \BibitemOpen
  \bibfield  {author} {\bibinfo {author} {\bibfnamefont {R.}~\bibnamefont
  {Suzuki}}, \bibinfo {author} {\bibfnamefont {T.}~\bibnamefont {Watanabe}}, \
  and\ \bibinfo {author} {\bibfnamefont {S.}~\bibnamefont {Ishihara}},\ }\href
  {\doibase 10.1103/PhysRevB.80.054410} {\bibfield  {journal} {\bibinfo
  {journal} {Phys. Rev. B}\ }\textbf {\bibinfo {volume} {80}},\ \bibinfo
  {pages} {054410} (\bibinfo {year} {2009})}\BibitemShut {NoStop}%
\bibitem [{\citenamefont {Kanamori}\ \emph {et~al.}(2011)\citenamefont
  {Kanamori}, \citenamefont {Matsueda},\ and\ \citenamefont
  {Ishihara}}]{Kanamori2011}%
  \BibitemOpen
  \bibfield  {author} {\bibinfo {author} {\bibfnamefont {Y.}~\bibnamefont
  {Kanamori}}, \bibinfo {author} {\bibfnamefont {H.}~\bibnamefont {Matsueda}},
  \ and\ \bibinfo {author} {\bibfnamefont {S.}~\bibnamefont {Ishihara}},\
  }\href {\doibase 10.1103/PhysRevLett.107.167403} {\bibfield  {journal}
  {\bibinfo  {journal} {Phys. Rev. Lett.}\ }\textbf {\bibinfo {volume} {107}},\
  \bibinfo {pages} {167403} (\bibinfo {year} {2011})}\BibitemShut {NoStop}%
\bibitem [{\citenamefont {Kanamori}\ \emph {et~al.}(2012)\citenamefont
  {Kanamori}, \citenamefont {Ohara},\ and\ \citenamefont
  {Ishihara}}]{Kanamori2012}%
  \BibitemOpen
  \bibfield  {author} {\bibinfo {author} {\bibfnamefont {Y.}~\bibnamefont
  {Kanamori}}, \bibinfo {author} {\bibfnamefont {J.}~\bibnamefont {Ohara}}, \
  and\ \bibinfo {author} {\bibfnamefont {S.}~\bibnamefont {Ishihara}},\ }\href
  {\doibase 10.1103/PhysRevB.86.045137} {\bibfield  {journal} {\bibinfo
  {journal} {Phys. Rev. B}\ }\textbf {\bibinfo {volume} {86}},\ \bibinfo
  {pages} {045137} (\bibinfo {year} {2012})}\BibitemShut {NoStop}%
\bibitem [{\citenamefont {Kune\ifmmode~\check{s}\else \v{s}\fi{}}\ and\
  \citenamefont {K\ifmmode~\check{r}\else \v{r}\fi{}\'apek}(2011)}]{Kunes2011}%
  \BibitemOpen
  \bibfield  {author} {\bibinfo {author} {\bibfnamefont {J.}~\bibnamefont
  {Kune\ifmmode~\check{s}\else \v{s}\fi{}}}\ and\ \bibinfo {author}
  {\bibfnamefont {V.}~\bibnamefont {K\ifmmode~\check{r}\else
  \v{r}\fi{}\'apek}},\ }\href {\doibase 10.1103/PhysRevLett.106.256401}
  {\bibfield  {journal} {\bibinfo  {journal} {Phys. Rev. Lett.}\ }\textbf
  {\bibinfo {volume} {106}},\ \bibinfo {pages} {256401} (\bibinfo {year}
  {2011})}\BibitemShut {NoStop}%
\bibitem [{\citenamefont {Kaneko}\ \emph {et~al.}(2012)\citenamefont {Kaneko},
  \citenamefont {Seki},\ and\ \citenamefont {Ohta}}]{Kaneko2012}%
  \BibitemOpen
  \bibfield  {author} {\bibinfo {author} {\bibfnamefont {T.}~\bibnamefont
  {Kaneko}}, \bibinfo {author} {\bibfnamefont {K.}~\bibnamefont {Seki}}, \ and\
  \bibinfo {author} {\bibfnamefont {Y.}~\bibnamefont {Ohta}},\ }\href {\doibase
  10.1103/PhysRevB.85.165135} {\bibfield  {journal} {\bibinfo  {journal} {Phys.
  Rev. B}\ }\textbf {\bibinfo {volume} {85}},\ \bibinfo {pages} {165135}
  (\bibinfo {year} {2012})}\BibitemShut {NoStop}%
\bibitem [{\citenamefont {Niyazi}\ \emph {et~al.}(2020)\citenamefont {Niyazi},
  \citenamefont {Geffroy},\ and\ \citenamefont {Kune\ifmmode~\check{s}\else
  \v{s}\fi{}}}]{Niyazi2020}%
  \BibitemOpen
  \bibfield  {author} {\bibinfo {author} {\bibfnamefont {A.}~\bibnamefont
  {Niyazi}}, \bibinfo {author} {\bibfnamefont {D.}~\bibnamefont {Geffroy}}, \
  and\ \bibinfo {author} {\bibfnamefont {J.}~\bibnamefont
  {Kune\ifmmode~\check{s}\else \v{s}\fi{}}},\ }\href {\doibase
  10.1103/PhysRevB.102.085159} {\bibfield  {journal} {\bibinfo  {journal}
  {Phys. Rev. B}\ }\textbf {\bibinfo {volume} {102}},\ \bibinfo {pages}
  {085159} (\bibinfo {year} {2020})}\BibitemShut {NoStop}%
\bibitem [{\citenamefont {Nasu}\ and\ \citenamefont {Naka}(2021)}]{Nasu2021}%
  \BibitemOpen
  \bibfield  {author} {\bibinfo {author} {\bibfnamefont {J.}~\bibnamefont
  {Nasu}}\ and\ \bibinfo {author} {\bibfnamefont {M.}~\bibnamefont {Naka}},\
  }\href {\doibase 10.1103/PhysRevB.103.L121104} {\bibfield  {journal}
  {\bibinfo  {journal} {Phys. Rev. B}\ }\textbf {\bibinfo {volume} {103}},\
  \bibinfo {pages} {L121104} (\bibinfo {year} {2021})}\BibitemShut {NoStop}%
\bibitem [{\citenamefont {Kune^^c5^^a1}(2015)}]{Kunes2015}%
  \BibitemOpen
  \bibfield  {author} {\bibinfo {author} {\bibfnamefont {J.}~\bibnamefont
  {Kune^^c5^^a1}},\ }\href {\doibase 10.1088/0953-8984/27/33/333201} {\bibfield
   {journal} {\bibinfo  {journal} {J. Phys.: Condens. Matter}\ }\textbf
  {\bibinfo {volume} {27}},\ \bibinfo {pages} {333201} (\bibinfo {year}
  {2015})}\BibitemShut {NoStop}%
\bibitem [{\citenamefont {Hoshino}\ and\ \citenamefont
  {Werner}(2016)}]{Hoshino2016}%
  \BibitemOpen
  \bibfield  {author} {\bibinfo {author} {\bibfnamefont {S.}~\bibnamefont
  {Hoshino}}\ and\ \bibinfo {author} {\bibfnamefont {P.}~\bibnamefont
  {Werner}},\ }\href {\doibase 10.1103/PhysRevB.93.155161} {\bibfield
  {journal} {\bibinfo  {journal} {Phys. Rev. B}\ }\textbf {\bibinfo {volume}
  {93}},\ \bibinfo {pages} {155161} (\bibinfo {year} {2016})}\BibitemShut
  {NoStop}%
\bibitem [{\citenamefont {Li}\ \emph {et~al.}(2020)\citenamefont {Li},
  \citenamefont {Otsuki}, \citenamefont {Naka},\ and\ \citenamefont
  {Ishihara}}]{Li2020}%
  \BibitemOpen
  \bibfield  {author} {\bibinfo {author} {\bibfnamefont {H.}~\bibnamefont
  {Li}}, \bibinfo {author} {\bibfnamefont {J.}~\bibnamefont {Otsuki}}, \bibinfo
  {author} {\bibfnamefont {M.}~\bibnamefont {Naka}}, \ and\ \bibinfo {author}
  {\bibfnamefont {S.}~\bibnamefont {Ishihara}},\ }\href {\doibase
  10.1103/PhysRevB.101.125117} {\bibfield  {journal} {\bibinfo  {journal}
  {Phys. Rev. B}\ }\textbf {\bibinfo {volume} {101}},\ \bibinfo {pages}
  {125117} (\bibinfo {year} {2020})}\BibitemShut {NoStop}%
\bibitem [{\citenamefont {Yamamoto}\ \emph {et~al.}(2020)\citenamefont
  {Yamamoto}, \citenamefont {Sugimoto},\ and\ \citenamefont
  {Ohta}}]{Yamamoto2020}%
  \BibitemOpen
  \bibfield  {author} {\bibinfo {author} {\bibfnamefont {S.}~\bibnamefont
  {Yamamoto}}, \bibinfo {author} {\bibfnamefont {K.}~\bibnamefont {Sugimoto}},
  \ and\ \bibinfo {author} {\bibfnamefont {Y.}~\bibnamefont {Ohta}},\ }\href
  {\doibase 10.1103/PhysRevB.101.174428} {\bibfield  {journal} {\bibinfo
  {journal} {Phys. Rev. B}\ }\textbf {\bibinfo {volume} {101}},\ \bibinfo
  {pages} {174428} (\bibinfo {year} {2020})}\BibitemShut {NoStop}%
\bibitem [{\citenamefont {Geffroy}\ \emph {et~al.}(2019)\citenamefont
  {Geffroy}, \citenamefont {Kaufmann}, \citenamefont {Hariki}, \citenamefont
  {Gunacker}, \citenamefont {Hausoel},\ and\ \citenamefont
  {Kune\ifmmode~\check{s}\else \v{s}\fi{}}}]{Geffroy2019}%
  \BibitemOpen
  \bibfield  {author} {\bibinfo {author} {\bibfnamefont {D.}~\bibnamefont
  {Geffroy}}, \bibinfo {author} {\bibfnamefont {J.}~\bibnamefont {Kaufmann}},
  \bibinfo {author} {\bibfnamefont {A.}~\bibnamefont {Hariki}}, \bibinfo
  {author} {\bibfnamefont {P.}~\bibnamefont {Gunacker}}, \bibinfo {author}
  {\bibfnamefont {A.}~\bibnamefont {Hausoel}}, \ and\ \bibinfo {author}
  {\bibfnamefont {J.}~\bibnamefont {Kune\ifmmode~\check{s}\else \v{s}\fi{}}},\
  }\href {\doibase 10.1103/PhysRevLett.122.127601} {\bibfield  {journal}
  {\bibinfo  {journal} {Phys. Rev. Lett.}\ }\textbf {\bibinfo {volume} {122}},\
  \bibinfo {pages} {127601} (\bibinfo {year} {2019})}\BibitemShut {NoStop}%
\bibitem [{\citenamefont {Mizokawa}\ and\ \citenamefont
  {Fujimori}(1996)}]{Mizokawa1996}%
  \BibitemOpen
  \bibfield  {author} {\bibinfo {author} {\bibfnamefont {T.}~\bibnamefont
  {Mizokawa}}\ and\ \bibinfo {author} {\bibfnamefont {A.}~\bibnamefont
  {Fujimori}},\ }\href {\doibase 10.1103/PhysRevB.54.5368} {\bibfield
  {journal} {\bibinfo  {journal} {Phys. Rev. B}\ }\textbf {\bibinfo {volume}
  {54}},\ \bibinfo {pages} {5368} (\bibinfo {year} {1996})}\BibitemShut
  {NoStop}%
\bibitem [{\citenamefont {Saitoh}\ \emph {et~al.}(1997)\citenamefont {Saitoh},
  \citenamefont {Mizokawa}, \citenamefont {Fujimori}, \citenamefont {Abbate},
  \citenamefont {Takeda},\ and\ \citenamefont {Takano}}]{Saitoh1997}%
  \BibitemOpen
  \bibfield  {author} {\bibinfo {author} {\bibfnamefont {T.}~\bibnamefont
  {Saitoh}}, \bibinfo {author} {\bibfnamefont {T.}~\bibnamefont {Mizokawa}},
  \bibinfo {author} {\bibfnamefont {A.}~\bibnamefont {Fujimori}}, \bibinfo
  {author} {\bibfnamefont {M.}~\bibnamefont {Abbate}}, \bibinfo {author}
  {\bibfnamefont {Y.}~\bibnamefont {Takeda}}, \ and\ \bibinfo {author}
  {\bibfnamefont {M.}~\bibnamefont {Takano}},\ }\href {\doibase
  10.1103/PhysRevB.55.4257} {\bibfield  {journal} {\bibinfo  {journal} {Phys.
  Rev. B}\ }\textbf {\bibinfo {volume} {55}},\ \bibinfo {pages} {4257}
  (\bibinfo {year} {1997})}\BibitemShut {NoStop}%
\bibitem [{\citenamefont {Hejtm\'anek}\ \emph {et~al.}(2010)\citenamefont
  {Hejtm\'anek}, \citenamefont {\ifmmode~\check{S}\else \v{S}\fi{}antav\'a},
  \citenamefont {Kn\'{\i}\ifmmode~\check{z}\else \v{z}\fi{}ek}, \citenamefont
  {Mary\ifmmode~\check{s}\else \v{s}\fi{}ko}, \citenamefont {Jir\'ak},
  \citenamefont {Naito}, \citenamefont {Sasaki},\ and\ \citenamefont
  {Fujishiro}}]{Hejtmanek2010}%
  \BibitemOpen
  \bibfield  {author} {\bibinfo {author} {\bibfnamefont {J.}~\bibnamefont
  {Hejtm\'anek}}, \bibinfo {author} {\bibfnamefont {E.}~\bibnamefont
  {\ifmmode~\check{S}\else \v{S}\fi{}antav\'a}}, \bibinfo {author}
  {\bibfnamefont {K.}~\bibnamefont {Kn\'{\i}\ifmmode~\check{z}\else
  \v{z}\fi{}ek}}, \bibinfo {author} {\bibfnamefont {M.}~\bibnamefont
  {Mary\ifmmode~\check{s}\else \v{s}\fi{}ko}}, \bibinfo {author} {\bibfnamefont
  {Z.}~\bibnamefont {Jir\'ak}}, \bibinfo {author} {\bibfnamefont
  {T.}~\bibnamefont {Naito}}, \bibinfo {author} {\bibfnamefont
  {H.}~\bibnamefont {Sasaki}}, \ and\ \bibinfo {author} {\bibfnamefont
  {H.}~\bibnamefont {Fujishiro}},\ }\href {\doibase 10.1103/PhysRevB.82.165107}
  {\bibfield  {journal} {\bibinfo  {journal} {Phys. Rev. B}\ }\textbf {\bibinfo
  {volume} {82}},\ \bibinfo {pages} {165107} (\bibinfo {year}
  {2010})}\BibitemShut {NoStop}%
\bibitem [{\citenamefont {Fujioka}\ \emph {et~al.}(2013)\citenamefont
  {Fujioka}, \citenamefont {Yamasaki}, \citenamefont {Nakao}, \citenamefont
  {Kumai}, \citenamefont {Murakami}, \citenamefont {Nakamura}, \citenamefont
  {Kawasaki},\ and\ \citenamefont {Tokura}}]{Fujioka2013}%
  \BibitemOpen
  \bibfield  {author} {\bibinfo {author} {\bibfnamefont {J.}~\bibnamefont
  {Fujioka}}, \bibinfo {author} {\bibfnamefont {Y.}~\bibnamefont {Yamasaki}},
  \bibinfo {author} {\bibfnamefont {H.}~\bibnamefont {Nakao}}, \bibinfo
  {author} {\bibfnamefont {R.}~\bibnamefont {Kumai}}, \bibinfo {author}
  {\bibfnamefont {Y.}~\bibnamefont {Murakami}}, \bibinfo {author}
  {\bibfnamefont {M.}~\bibnamefont {Nakamura}}, \bibinfo {author}
  {\bibfnamefont {M.}~\bibnamefont {Kawasaki}}, \ and\ \bibinfo {author}
  {\bibfnamefont {Y.}~\bibnamefont {Tokura}},\ }\href {\doibase
  10.1103/PhysRevLett.111.027206} {\bibfield  {journal} {\bibinfo  {journal}
  {Phys. Rev. Lett.}\ }\textbf {\bibinfo {volume} {111}},\ \bibinfo {pages}
  {027206} (\bibinfo {year} {2013})}\BibitemShut {NoStop}%
\bibitem [{\citenamefont {Fujioka}\ \emph {et~al.}(2015)\citenamefont
  {Fujioka}, \citenamefont {Yamasaki}, \citenamefont {Doi}, \citenamefont
  {Nakao}, \citenamefont {Kumai}, \citenamefont {Murakami}, \citenamefont
  {Nakamura}, \citenamefont {Kawasaki}, \citenamefont {Arima},\ and\
  \citenamefont {Tokura}}]{Fujioka2015}%
  \BibitemOpen
  \bibfield  {author} {\bibinfo {author} {\bibfnamefont {J.}~\bibnamefont
  {Fujioka}}, \bibinfo {author} {\bibfnamefont {Y.}~\bibnamefont {Yamasaki}},
  \bibinfo {author} {\bibfnamefont {A.}~\bibnamefont {Doi}}, \bibinfo {author}
  {\bibfnamefont {H.}~\bibnamefont {Nakao}}, \bibinfo {author} {\bibfnamefont
  {R.}~\bibnamefont {Kumai}}, \bibinfo {author} {\bibfnamefont
  {Y.}~\bibnamefont {Murakami}}, \bibinfo {author} {\bibfnamefont
  {M.}~\bibnamefont {Nakamura}}, \bibinfo {author} {\bibfnamefont
  {M.}~\bibnamefont {Kawasaki}}, \bibinfo {author} {\bibfnamefont
  {T.}~\bibnamefont {Arima}}, \ and\ \bibinfo {author} {\bibfnamefont
  {Y.}~\bibnamefont {Tokura}},\ }\href {\doibase 10.1103/PhysRevB.92.195115}
  {\bibfield  {journal} {\bibinfo  {journal} {Phys. Rev. B}\ }\textbf {\bibinfo
  {volume} {92}},\ \bibinfo {pages} {195115} (\bibinfo {year}
  {2015})}\BibitemShut {NoStop}%
\bibitem [{\citenamefont {Yamasaki}\ \emph {et~al.}(2016)\citenamefont
  {Yamasaki}, \citenamefont {Fujioka}, \citenamefont {Nakao}, \citenamefont
  {Okamoto}, \citenamefont {Sudayama}, \citenamefont {Murakami}, \citenamefont
  {Nakamura}, \citenamefont {Kawasaki}, \citenamefont {Arima},\ and\
  \citenamefont {Tokura}}]{Yamasaki2016}%
  \BibitemOpen
  \bibfield  {author} {\bibinfo {author} {\bibfnamefont {Y.}~\bibnamefont
  {Yamasaki}}, \bibinfo {author} {\bibfnamefont {J.}~\bibnamefont {Fujioka}},
  \bibinfo {author} {\bibfnamefont {H.}~\bibnamefont {Nakao}}, \bibinfo
  {author} {\bibfnamefont {J.}~\bibnamefont {Okamoto}}, \bibinfo {author}
  {\bibfnamefont {T.}~\bibnamefont {Sudayama}}, \bibinfo {author}
  {\bibfnamefont {Y.}~\bibnamefont {Murakami}}, \bibinfo {author}
  {\bibfnamefont {M.}~\bibnamefont {Nakamura}}, \bibinfo {author}
  {\bibfnamefont {M.}~\bibnamefont {Kawasaki}}, \bibinfo {author}
  {\bibfnamefont {T.}~\bibnamefont {Arima}}, \ and\ \bibinfo {author}
  {\bibfnamefont {Y.}~\bibnamefont {Tokura}},\ }\href {\doibase
  10.7566/JPSJ.85.023704} {\bibfield  {journal} {\bibinfo  {journal} {J. Phys.
  Soc. Jpn.}\ }\textbf {\bibinfo {volume} {85}},\ \bibinfo {pages} {023704}
  (\bibinfo {year} {2016})}\BibitemShut {NoStop}%
\bibitem [{\citenamefont {Yokoyama}\ \emph {et~al.}(2018)\citenamefont
  {Yokoyama}, \citenamefont {Yamasaki}, \citenamefont {Taguchi}, \citenamefont
  {Hirata}, \citenamefont {Takubo}, \citenamefont {Miyawaki}, \citenamefont
  {Harada}, \citenamefont {Asakura}, \citenamefont {Fujioka}, \citenamefont
  {Nakamura}, \citenamefont {Daimon}, \citenamefont {Kawasaki}, \citenamefont
  {Tokura},\ and\ \citenamefont {Wadati}}]{Yokoyama2018}%
  \BibitemOpen
  \bibfield  {author} {\bibinfo {author} {\bibfnamefont {Y.}~\bibnamefont
  {Yokoyama}}, \bibinfo {author} {\bibfnamefont {Y.}~\bibnamefont {Yamasaki}},
  \bibinfo {author} {\bibfnamefont {M.}~\bibnamefont {Taguchi}}, \bibinfo
  {author} {\bibfnamefont {Y.}~\bibnamefont {Hirata}}, \bibinfo {author}
  {\bibfnamefont {K.}~\bibnamefont {Takubo}}, \bibinfo {author} {\bibfnamefont
  {J.}~\bibnamefont {Miyawaki}}, \bibinfo {author} {\bibfnamefont
  {Y.}~\bibnamefont {Harada}}, \bibinfo {author} {\bibfnamefont
  {D.}~\bibnamefont {Asakura}}, \bibinfo {author} {\bibfnamefont
  {J.}~\bibnamefont {Fujioka}}, \bibinfo {author} {\bibfnamefont
  {M.}~\bibnamefont {Nakamura}}, \bibinfo {author} {\bibfnamefont
  {H.}~\bibnamefont {Daimon}}, \bibinfo {author} {\bibfnamefont
  {M.}~\bibnamefont {Kawasaki}}, \bibinfo {author} {\bibfnamefont
  {Y.}~\bibnamefont {Tokura}}, \ and\ \bibinfo {author} {\bibfnamefont
  {H.}~\bibnamefont {Wadati}},\ }\href {\doibase
  10.1103/PhysRevLett.120.206402} {\bibfield  {journal} {\bibinfo  {journal}
  {Phys. Rev. Lett.}\ }\textbf {\bibinfo {volume} {120}},\ \bibinfo {pages}
  {206402} (\bibinfo {year} {2018})}\BibitemShut {NoStop}%
\end{thebibliography}%

\end{document}